\documentclass[a4paper,11pt]{article}
\usepackage{jheppub}

\usepackage{ifpdf}
\usepackage{xcolor}
\usepackage{amsmath,amssymb,amsfonts,bm,amscd}
\usepackage{amscd}
\usepackage{mathtools}
\usepackage{graphicx, wrapfig}
\usepackage{multirow}
\usepackage{verbatim}
\usepackage{appendix}
\usepackage{fancybox}
\usepackage{slashed}
\usepackage{extarrows}
\usepackage{braket}
\usepackage{url}
\usepackage{float}
\usepackage{bbold}

\usepackage{mathtools}
\usepackage{mathdots}
\usepackage{amsfonts}
\usepackage{amssymb}
\usepackage{tikz}
\usepackage{tikz-cd}
\usetikzlibrary{patterns,cd,decorations.pathmorphing,positioning, shapes.geometric}
\usepackage{todonotes}
\usepackage{hyperref}

\newcommand{\field}[1]{\mathbb{#1}}
\newcommand{\CC}{\field{C}}

\newcommand{\ZZ}{\field{Z}}

\newcommand{\NN}{\field{N}}

\newcommand{\N}{\mathcal{N}}
\newcommand{\sheaf}[1]{\mathcal{#1}}
\renewcommand{\O}{\sheaf{O}}

\newcommand{\geom}[1]{\mathfrak{#1}}

\newcommand{\T}{\geom{T}}

\newcommand{\id}{\mathrm{id}}
\newcommand{\II}{\mathbb{I}}
\newcommand{\orb}[1]{\widetilde{#1}}

\newcommand{\glsm}[1]{\widehat{#1}}
\newcommand{\red}{\text{red}}
\newcommand{\LG}{\text{LG}}

\renewcommand{\d}{\text{d}}
\newcommand{\bZ}{\mathbb{Z}}
\newcommand{\bC}{\mathbb{C}}

\renewcommand{\d}{\text{d}}

\DeclareMathOperator{\cone}{Cone}

\DeclareMathOperator{\coker}{coker}

\tikzset{LG/.style={rectangle, fill=blue!10, minimum height=2cm, minimum width=2cm}}
\tikzset{GLSM/.style={rectangle, fill=blue!25, minimum height=2cm, minimum width=2cm}}
\tikzset{geom/.style={rectangle, 
minimum height=2cm, minimum width=2cm,
pattern=north east lines, 
pattern color=blue!30}}
\usepackage{tikz}
\usetikzlibrary{decorations.markings}
\usetikzlibrary{decorations.pathreplacing,angles,quotes}
\usetikzlibrary{decorations.pathmorphing,shapes}
\newcounter{sarrow}

\definecolor{WScolor}{RGB}{191,191,255}
\definecolor{WScolor light}{RGB}{224,224,255}
\definecolor{DefectColor}{RGB}{0,0,128}
\tikzset{
	defect/.style={color=DefectColor, line width=1.5},
	arrow position/.style={postaction={decorate,decoration={
		markings,
		mark=at position #1 with {\arrow{>}}
	}}},
	opp arrow position/.style={postaction={decorate,decoration={
		markings,
		mark=at position #1 with {\arrow{<}}
	}}},
	defect node/.style={circle,inner sep=1.5pt,fill, DefectColor},
	snake it/.style={
		-stealth,
		decoration={
			snake, 
    		amplitude = .4mm,
    		segment length = 2mm,
    		post length=0.9mm
    	},
    	decorate
    }
}
\tikzcdset{scale cd/.style={every label/.append style={scale=#1},
    cells={nodes={scale=#1}}}}

\title{Defects and phase transitions to geometric phases of abelian GLSMs}

\author[1]{Ilka Brunner,}
\author[1]{Lukas Krumpeck,}
\author[3]{Daniel Roggenkamp}
\affiliation[1]{Arnold Sommerfeld Center, Ludwig-Maximilians-Universit\"at,\\
Theresienstra{\ss}e 37, 80333 M\"unchen, Germany}
\affiliation[3]{Institut f\"ur Mathematik, Universit\"at Mannheim,\\
B6, 26, 68131 Mannheim, Germany}

\abstract{We consider gauged linear sigma models with gauge group $U(1)$ that exhibit a geometric as well as a Landau Ginzburg phase. We construct defects that implement the transport of D-branes  from the Landau-Ginzburg phase to the geometric phase. Through their fusion with boundary conditions these defects in particular provide functors between the respective D-brane categories. The latter map (equivariant) matrix factorizations to coherent sheaves and can be formulated explicitly in terms of complexes of matrix factorizations.}

\preprint{LMU-ASC 27/21}

\begin{document}

\cornersize{1}
\maketitle
\section{Introduction}

We consider two-dimensional gauged linear sigma models with $(2,2)$ supersymmetry, $U(1)$ gauge group, charged matter multiplets and a superpotential. Depending on the complexified FI parameter, such models can in general exhibit different phases, characterized by a partial or total breaking of the gauge symmetry \cite{Witten-phases}.

In geometric phases, the gauge symmetry is typically completely broken and the models reduce to non-linear sigma models, whose target manifolds are projective hypersurfaces.
In Landau-Ginzburg phases, the gauge symmetry is typically broken to a finite subgroup and the model reduces to a Landau-Ginzburg orbifold theory. 

It has been proposed in \cite{BR-new} that transitions between different phases of such gauged linear sigma models can be described by means of defect lines. Via their fusion with boundaries,  these transition defects give rise to functors between  the respective D-brane categories. In \cite{BR-new} transition defects and the respective functors have been explicitly described for transitions between different Landau-Ginzburg phases. 
The aim of the present paper is to explicitly construct defect lines implementing the transition between different kinds of phases, more precisely from Landau-Ginzburg to  geometric phases. 

We focus on the degrees of freedom surviving a topological B-twist, decoupling the gauge sector. In this setting, it is well known how to describe the categories of D-branes preserving B-type supersymmetry; in geometric phases it is given by the derived category of coherent sheaves  on the target space, whereas at  Landau-Ginzburg points 
it is the category of equivariant matrix factorizations of the respective superpotential.
The transport of D-branes from one phase to the other thus  has to mediate between these different categories.

For the concrete construction, we follow the strategy proposed in \cite{BR-new}. The starting point is the trivial defect line of the gauged linear sigma model, which is given in terms of a $U(1)$ equivariant matrix factorization. To turn this into a defect that connects the two phases, one pushes the gauged linear sigma model on one side of the trivial defect to the geometric phase, and the one on the other side to the Landau-Ginzburg phase.  
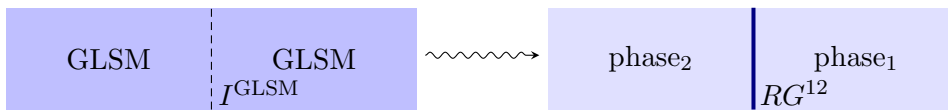
\begin{equation}\nonumber \tikz[baseline=10,scale=0.9]{
	\fill[WScolor] (-2,-.25) rectangle (4,1.25);
	\draw[densely dashed] (1,-.25) -- (1,1.25);
	\node at (1.7,0) {$I^\text{GLSM}$};
	\node at (-.5,.5) {GLSM};
	\node at (2.5,.5) {GLSM};
}\;\tikz[baseline=-4]{
	\path (0,0) edge[snake it] (1.5,0);
}\;\tikz[baseline=10,scale=0.9]{
	\fill[WScolor light] (-2,-.25) rectangle (1,1.25);
	\fill[WScolor light] (1,-.25) rectangle (4,1.25);
	\draw[defect] (1,-.25) -- (1,1.25);
	\node at (-.5,.5) {phase$_2$};
	\node at (2.5,.5) {phase$_1$};
	\node at (1.6,0) {$RG^{12}$};
}\end{equation}

In the phases, specific field configurations are excluded and the transition to a phase  takes this into account. For example, in Landau-Ginzburg phases certain fields obtain a non-vanishing vacuum expectation value, and pushing the GLSM to a Landau-Ginzburg phase involves setting the field to its vev. We implement this explicitly  when constructing the defect. 

In fact, in general there are different paths connecting two phases.  This means that in general there is more than one transition defect connecting a pair of phases, and additional data is required to uniquely specify it. This additional data appeared in \cite{BR-new} in terms of a choice of cutoff parameter which has to be introduced when pushing the GLSM into the phases on the two sides of the identity defect.

It turns out that the transition defects $RG^{12}$ between two phases of a GLSM always factorize over the GLSM, i.e.~$RG^{12}$ can be obtained as the fusion of a defect $T^1$ from phase$_1$ to the GLSM and a defect $R^2$ from the GLSM to phase$_2$:
$$
RG^{12}\cong R^2\otimes T^1 \ .
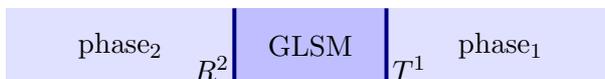
$$
\[\tikz[baseline=10]{
	\fill[WScolor light] (-2,0) rectangle (1,1);
	\fill[WScolor] (1,0) rectangle (3,1);
	\fill[WScolor light] (3,0) rectangle (6,1);
	\draw[defect] (1,0) -- (1,1);
	\draw[defect] (3,0) -- (3,1);
	\node at (4.5,.5) {phase$_1$};
	\node at (2,.5) {GLSM};
	\node at (-.5,.5) {$\text{phase$_2$}$};
	\node at (.7,.2) {$R^2$};
	\node at (3.3,.2) {$T^1$};
}\]
The defect $T^1$ embeds phase$_1$ into the GLSM. In particular, it 
lifts D-branes from that phase to the GLSM, thereby specifying a subcategory of the GLSM branes into which the D-branes of phase$_1$ are embedded. Transition via different paths leads to different transition defects and different subcategories into which phase$_1$ D-branes are mapped. 
In the language of \cite{HHP_08}, the lifted  D-branes are called grade restricted and the choice of path corresponds to a choice of ``window" of allowed representation labels in \cite{HHP_08,hori_exact_2013,Knapp:2016rec,RlFC_brane-transport_18}. In our construction, the grade restriction rule arises automatically as a consequence of supersymmetry and the rigidity of the defect construction. 

The action of defects on D-branes is implemented very concretely by merging the line defects with the boundary conditions specified by the D-branes.

\[\begin{aligned}\tikz[baseline=17,scale=0.8]{
		\fill[WScolor light] (-1,.25) rectangle (2,1.25);
		\draw[defect] (2,.25) -- (2,1.25);
		\node at (1.7,.45) {$B$};
		\node at (.5,.75) {LG};
	}\quad&\longmapsto\quad\tikz[baseline=17, scale=0.8]{
		\fill[WScolor light] (-2,.25) rectangle (1,1.25);
		\fill[WScolor light] (1,.25) rectangle (4,1.25);
		\draw[defect] (1,.25) -- (1,1.25);
		\draw[defect] (4,.25) -- (4,1.25);
		\node at (.7,.45) {$D$};
		\node at (-.5,.75) {geom.};
		\node at (2.5,.75) {LG};
		\node at (3.7,.45) {$B$};
	}\phantom{*}&\tikz[baseline=-2]{
	\path (0,0) edge[snake it] (0.7,0);
}\;\phantom{**}
	\tikz[baseline=17,scale=0.8]{
		\fill[WScolor light] (-1,.25) rectangle (2.5,1.25);
		\draw[defect] (2.5,.25) -- (2.5,1.25);
		\node at (1.7,.45) {$D\otimes B$};
		\node at (0.4,.75) {geom.};
	}
\end{aligned} \]

This action is smooth and well defined in the topological subsector and explicitly computable using the techniques of \cite{BR-LG-defects,KR_18,BR-new}. 

Since our arguments only make use of the topological subsector, they are unaffected by potential anomalies of the R-symmetry. Thus, they can be applied to both non-anomalous and anomalous models. 
For non-anomalous models where the axial R-symmetry is preserved at the quantum level, the target manifold is Calabi-Yau. Upon flowing to the IR, the parameter of the gauged linear sigma model becomes a K\"ahler modulus of a family of Calabi-Yau manifolds. The Landau-Ginzburg phase can be regarded as the stringy regime of the non-linear sigma model on the Calabi-Yau manifold. The transition between LG and geometric phase can be thought of as a deformation of the non-linear sigma model from the stringy small volume to the large volume regime.

In the anomalous case, it is still possible to find different phases in a single gauged linear sigma model. While the RG flow drives the model to a particular IR phase, a second phase can often be embedded by tuning parameters. We refer to it as the UV phase. In the discussion of this paper, the LG phase lies at the UV (in case the model is anomalous), and we consider flows starting there and ending in a geometric phase, which is the proper IR limit of the GLSM. The transition can therefore be thought of as a relevant flow.
Under this flow, some D-branes decouple to the Coulomb branch.  On the level of the topological subsector, they disappear from the theory and are projected out. The grade restriction rule for the anomalous case has been discussed in \cite{hori_exact_2013,RlFC_brane-transport_18}.

This paper is organized as follows. In section \ref{sec2} we explain how one can make the ideas outlined above concrete. To do so, we briefly recall how to describe B-type supersymmetric boundary conditions and defects in the  GLSM and its LG phases in terms of matrix factorizations. In particular, we recall the construction of the identity defect in GLSMs given in \cite{BR-new}.
In section~\ref{sec:Transition} we spell out the 
 construction of transition defects between GLSMs and LG phases by 
pushing down the GLSM on one side of the identity defect to a LG phase.  
Next, we explain how to push the GLSM on the other side of the resulting defect to the geometric phase.
This yields a transition defect from the LG to the geometric phase. It 
 shares features of matrix factorizations and complexes and can be regarded as a complex of matrix factorizations or in terms of nested cones. We discuss this in section \ref{sec:LGgeom}. Finally, in section \ref{sec:LGgeomaction} we explain how to fuse the transition defects with boundary conditions. This describes how D-branes behave under the transition.
 
Section \ref{sec:example} contains the concrete calculations for the example of the GLSM with superpotential $W=PG(X_1,\ldots,X_N)$, where $G$ is a homogeneous polynomial in the $X_i$. In the geometric phase, these models reduce to non-linear sigma models on the projective hypersurface $\{G(X_1,\ldots,X_N)=0\}\subset\mathbb{P}^{N-1}$, whereas at the LG point, they 
are effectively described by LG models with chiral superfields $X_i$ and superpotential $G(X_1,\ldots,X_N)$. 
Here, we start out by giving concrete formulas for the transition defects between GLSM and LG phase. 
We show that the associated functors map the D-branes of the LG phase to 
grade restricted subcategories of GLSM D-brane category in the sense of \cite{HHP_08}.  Subsequently, we construct the transition defects to the geometric phase and compute their action on D-branes. For the Calabi-Yau case, our procedure reproduces known results from \cite{HHP_08} in a novel way.

\section{Transition defects between different phases of GLSMs} \label{sec2}

\subsection{Defects in GLSMs and Landau-Ginzburg models}\label{sec:Defects_GLSM_LG}

In the B-twisted sector, the GLSM is effectively described by the matter fields subject to the superpotential. In this setting, the GLSM can be regarded as an equivariant Landau-Ginzburg model. D-branes as well as defects are then described in terms of (equivariant) matrix factorizations of the superpotential. 
To establish our notation, let us briefly recall that a matrix factorization of a polynomial $W$ over a polynomial ring $S_{(X)(Y)}:=\bC[X_1,\ldots,X_n, Y_1, \ldots Y_m]$ is given by
\begin{equation*}
P:P_1
\\ 
	\tikz[baseline=0]{
		\node at (0,.5) {$p_1$};
		\draw[arrow position = 1] (-1.5,.2) -- (1.5,.2);
		\draw[arrow position = 1] (1.5,0) -- (-1.5,0);
		\node at (0,-.4) {$p_0$};
	}\, P_0 \,
\end{equation*}
where   $P=P_0\oplus P_1$ is a $\bZ_2$-graded free module  over $S_{(X)(Y)}$ and 
\[
	\d_P=\begin{pmatrix}0 & p_1 \\ p_0 & 0\end{pmatrix}\quad
\]
is an odd endomorphism of $P$ such that $\d_P^2 = W\cdot \id_P$.

Defects separating a Landau-Ginzburg model with chiral fields $X_1,\ldots,X_n$ and superpotential $W_1(X_1,\ldots,X_n)$ from one with chiral fields $Y_1,\ldots, Y_m$ and superpotential $W_2(Y_1,\ldots,Y_m)$
are described by matrix factorizations of the difference $W=W_1(X_i)- W_2(Y_i)$ of the two superpotentials.

For our applications, we need to consider $G$-equivariant models, where $G$ is a finite or continuous abelian group. Most relevant for the example is the case $G=U(1)$ for the GLSM, and $G= {\mathbb Z}_d$ for the LG phase. In order to describe defects between different models, we need to admit different groups 
$G_1$ and $G_2$ on the two sides of the defect. Accordingly, the modules $P_0$ and $P_1$ must carry $G_1 \times G_2$-representations and the maps $p_1$ and $p_0$ are subject to an equivariance condition. We refer to appendix A of \cite{BR-new} for further details.  

Instead of dealing with the matrix factorizations themselves, it is sometimes useful to consider associated modules
$M_P$ over the  quotient ring $C=S_{(X)(Y)}/(W)$, admitting a free resolution which after finitely many steps turns into a two-periodic one defined by the matrix factorization, i.e.
\begin{equation*}
\ldots\stackrel{p_0}{\longrightarrow} P_1\otimes_SC\stackrel{p_1}{\longrightarrow} P_0\otimes_SC\stackrel{p_0}{\longrightarrow} P_1\otimes_SC\stackrel{p_1}{\longrightarrow} P_0\otimes_SC
\stackrel{\varphi_{m+1}}{\longrightarrow} M_P^m\stackrel{\varphi_m}{\longrightarrow} 
\ldots \stackrel{\varphi_1}{\longrightarrow} M_P^0=M_P
\rightarrow 0 \ .
\end{equation*} 
An example is 
\begin{equation}\label{eq:module}
M_P=\text{coker}(p_1:P_1\otimes_SC\rightarrow P_0\otimes_SC)\,,
\end{equation}
which has a free resolution which is two-periodic from the start \cite{Eisenbud}. 
Importantly, isomorphisms between modules $M_P$ and $M_Q$ associated to matrix factorizations $P$ and $Q$ of the same polynomial $W$ lift to the resolutions and give rise to isomorphisms of the respective matrix factorizations. 
This carries over to the case of equivariant matrix factorizations, and we will make excessive use of it.

As our strategy and notation is taken from \cite{BR-new}, we refer to that paper for further explanations on the description of defects and their fusion by means of (equivariant) matrix factorizations. A collection of details and more references can be found in the appendix of \cite{BR-new}.

\subsection{The identity defect in GLSMs and LG phases}\label{sec:iddef}

Most important for the current paper is the existence of identity defects in abelian GLSMs, which have been constructed in \cite{BR-new}. 
Consider a GLSM with a number of $n$ superfields $X_i$ with charges $Q_i$ under a $U(1)$ gauge group, and superpotential $W(X_i)$. The identity defect in this GLSM can now be described by a specific $U(1)\times U(1)$-equivariant matrix factorization of the difference of superpotentials $W(X_i)-W(Y_i)$. ($Y_i$ denote the chiral superfields on the other side of the defect. The fields $X_i$ are charged under the first $U(1)$-factor, and the $Y_i$ under the second.)

The identity defect in this GLSM can now be constructed by introducing a pair of bosonic defect fields $\alpha, \alpha^{-1}$  of $U(1)\times U(1)$ charge $(-1,1)$ \cite{BR-new}. It is given by 
 the $U(1)\times U(1)$-equivariant matrix factorization 
\begin{equation}\label{eq:continuous-id}
I=I_0 \oplus I_1 = S_{(X),(Y)}\otimes\Lambda(V)\otimes\bC[\alpha,\alpha^{-1}]/(\alpha\alpha^{-1}-1)
\end{equation}
with differential
\[\begin{aligned}
	&\d_I = \sum_{i=1}^n\left[ (X_i-\alpha^{Q_i} Y_i)\cdot \theta^*_i + \partial^{X,\alpha Y}_{i}W\cdot \theta_i \right]\,, \quad\\
	&\partial^{X,\alpha Y}_iW = \frac{W(\alpha^{Q_1}Y_1,..., \alpha^{Q_{i-1}}Y_{i-1}, X_i, ..., X_n) - W(\alpha^{Q_1}Y_1, ..., \alpha^{Q_i}Y_{i}, X_{i+1}, ..., X_n)}{X_i-\alpha^{Q_i} Y_{i}}\,.
\end{aligned}\]
Here, $V$ is a vector space with basis $\theta_1, \dots, \theta_n$. $\Lambda(V)$ is its exterior algebra,  and
$\theta_1^*,\ldots,\theta_n^*$ denotes the dual basis of $V^*$ (i.e $\theta_i^*(\theta_j)= \delta_{ij}$). 
Note that this matrix factorization is of infinite rank over the initial polynomial ring. This is because we allowed for any powers of the new fields $\alpha$ and $\alpha^{-1}$. 

To this matrix factorization we can associate the $C_{(X)(Y)}:=S_{(X)(Y)}/(W(X_i)-W(Y_i))$-module 
\begin{equation}\label{eq:id-module}
M_I=C_{(X)(Y)}[\alpha,\alpha^{-1}] / (X_i-\alpha^{Q_i} Y_i,\alpha\alpha^{-1}-1 ) \ ,
\end{equation}
whose Koszul resolution turns into the two-periodic resolution defined by the matrix factorization $I$. 
It has been checked in \cite{BR-new} that the defect associated to $I$ indeed acts trivially on any equivariant  matrix factorization.

To spell out the identity defect of a Landau-Ginzburg model with finite orbifold group $\bZ_d$, one simply
 adds a bosonic field on which one imposes $\alpha^d=1$ and regards the module
\begin{equation}\label{eq:LG-id}
M_{I^{\text{LG}}}=S_{(X),(Y)}\otimes\Lambda(V)\otimes\bC[\alpha]/(\alpha^d-1)
\end{equation}
instead of (\ref{eq:continuous-id}). This corresponds to the finite rank matrix factorization known to describe the identity defect in the LG orbifold model \cite{BR_LG-orbifold_08}.

\subsection{The models}\label{sec:Themodels}

The methods presented in this paper are very general. To make the construction of transition defects very explicit, we will focus on a specific class of GLSMs. The models have $U(1)$ gauge group, $N$ charged matter 
fields $X_i$ of charge $Q_i=1$, another charged matter field $P$ of charge $-d$ and a superpotential
of the form
$$
W(X_i)=P G(X_i)\,.
$$
$G(X_i)$ is a homogeneous polynomial in the $X_i$ of degree $d$. This ensures that the superpotential is gauge invariant. 
A consistent assignment of $R$-charges in the GLSM is to give R-charge $0$, respectively $2$ to $X_i$ and $P$.

It is well known that these models exhibit a Landau-Ginzburg as well as a geometric phase. These have been discussed in \cite{Witten-phases}. On a classical level they can be found by minimizing the potential for the scalars. In the LG phase, $P$ acquires a vacuum expectation value, breaking the gauge symmetry to $\ZZ_d$. 
In this phase, the theory is effectively described by a $\ZZ_d$-orbifold of a Landau-Ginzburg model with chiral superfields $X_i$ and superpotential $W(X_i)$. 
In the geometric phase, the configuration $X_i=0$ for all $i$ is not allowed and must be excluded. The gauge symmetry is completely broken and the model is described by a non-linear sigma model whose target space is the 
hypersurface in projective space parametrized by the $X_i$ defined by the equation 
$G(X_i)=0$.

If $d=N$, the R-symmetry is preserved at the quantum level. The Fayet-Iliopoulos parameter is exactly marginal, and LG and geometric phase can be connected by a marginal deformation. (In that case the target space is a Calabi-Yau manifold.)

If $d\neq N$, on the other hand, R-symmetry is broken, the Fayet-Illiopoulos parameter is a running coupling constant and the theory flows to one of the phases in the IR (the geometric phase for $d>N$ and the LG phase for $d<N$). The other phase can then be embedded as a UV phase into the same GLSM. In this case, some vacua will decouple along the flow from UV to IR. 

While the construction of the transition defects from the LG to the geometric phase is completely general,
we will be mostly interested in the case where the geometric phase lies in the IR ($d\geq N$). In that case the transition defects have a nice interpretation as the defects associated to the respective RG flow \cite{BR_LG-orbifold_08}.

\subsection{Transition defects between GLSMs and LG phases}\label{sec:Transition}

Transition defects between abelian GLSMs and LG phases have been constructed in \cite{BR-new}. We will give a brief sketch of the construction here. Starting point is the 
identity defect in the GLSM described in section~\ref{sec:iddef}.
It is represented by a $U(1)\times U(1)$-equivariant matrix factorization $I$ of $PG(X_i)-QG(Y_i)$. Here $(P,X_1,\ldots,X_N)$ denote the fields on the left of the defect and $(Q,Y_1,\ldots,Y_N)$ the ones on the right. For the construction of the transition defects, it is more convenient to work with the associated modules
\begin{equation}
M_I=C_{(X,P)(Y,Q)}[\alpha,\alpha^{-1}] / (P-\alpha^{Q_P} Q,X_i-\alpha^{Q_i} Y_i,\alpha\alpha^{-1}-1 ) \ ,
\end{equation}
c.f.~equation (\ref{eq:id-module}). Here $C_{(X,P)(Y,Q)}=\CC[X_1,\ldots,X_N,P,Y_1,\ldots,Y_N,Q]/(PG(X_i)-QG(Y_i))$. In order to obtain a transition defect, one pushes the GLSM on one side of the defect into the LG phase. 
This involves giving vacuum expectation values to the fields $P$, resp.~$Q$. 
On the level of the modules $M_I$ one has to set the respective variable to $1$.

Pushing the GLSM into the LG phase on both sides of the identity defect, for instance, requires to set $P=Q=1$ in $M_I$. This automatically imposes $\alpha^{Q_P}=1$ and implements the correct truncation of the above infinite dimensional module to a finite dimensional one. Indeed, one arrives at the module (\ref{eq:LG-id}) associated to the identity defect in the Landau-Ginzburg orbifold models. 

To obtain a transition defect from the LG phase to the GLSM one only has to push the GLSM on the right of the identity defect to the LG phase, i.e. one only sets $Q=1$ in $M_I$. The resulting module is a priori of infinite rank, and as proposed in \cite{BR-new} one has to introduce a cutoff $N$
for the highest power of $\alpha$ that appears in the module. This yields finite rank modules
\begin{equation}\label{eq:T_module}
M_I^{\text{GLSM}, \, \LG}(N)=\alpha^NC_{(X,P)(\dot,Y)} [\alpha^{-1}]/ (P-\alpha^{Q_P} , X_i-\alpha^{Q_i} Y_i) 
\alpha^N C_{(X,P)(\dot,Y)} [\alpha^{-1}]\ ,
\end{equation}
which depends on the 
cutoff parameter $N$.
These modules correspond to 
concrete matrix factorizations of the potential
\begin{equation}\label{eq:SP_T}
	W(P,X_i,Y_i) = P\cdot G(X_i) - G(Y_i) 
\end{equation}
which we denote by
\begin{equation}\label{eq:transition-mf}
	\begin{tikzcd}
		T_b : \; T_1 \rar[shift left]{t_1} & \lar[shift left]{t_0} T_0 .
	\end{tikzcd}
\end{equation}
For later convenience, we have labelled them by $b:=N-d+1$ instead of $N$. 
These matrix factorizations indeed describe the transition defects between LG phase and the GLSM \cite{BR-new}, where $N$, resp. $b$ parametrizes the choice of a path. 

In the simple example studied in this paper, there is an alternative way to obtain the transition defects, which we will employ in the discussion of the example in section~\ref{sec:example}.
Instead off starting with the identity defect in the GLSM and pushing the model on the left side to the LG phase,
one starts with the identity defect of the Landau-Ginzburg phase and then, following the prescriptions given in \cite{BJR-Monodromies} (which in turn follows \cite{HHP_08}), lifting the model on its left side to the GLSM.
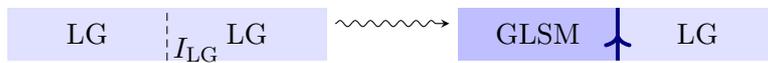
\begin{equation}\nonumber\tikz[baseline=10,scale=0.7]{
	\fill[WScolor light] (-2,0) rectangle (4,1);
	\draw[densely dashed] (1,0) -- (1,1);
	\node at (1.55,.2) {$I_\text{LG}$};
	\node at (-.5,.5) {LG};
	\node at (2.5,.5) {LG};
}\;\tikz[baseline=-4]{
	\path (0,0) edge[snake it] (1.5,0);
}\;\tikz[baseline=10,scale=0.7]{
	\fill[WScolor ] (-2,0) rectangle (1,1);
	\fill[WScolor light] (1,0) rectangle (4,1);
	\draw[defect, arrow position=.5] (1,0) -- (1,1);
	\node at (-.5,.5) {GLSM};
	\node at (2.5,.5) {LG};
	\node at (1.3,.2) {$$};
}\end{equation}
On the level of defects this involves placing suitable powers of $P$ into entries of the matrix factorization, 
so as to produce a matrix factorization of $PG(X_i)-G(Y_i)$ from the identity matrix factorization of the LG model, which is a matrix factorization of $G(X_i)-G(Y_i)$. Since $P$ appears linearly in the superpotential of our model, (up to equivalence) there is only one choice. 
However, the resulting matrix factorization has to be $U(1)\times\ZZ_d$-equivariant. Thus, one has to lift $\ZZ_d$-representations of the original LG-identity matrix factorizations to $U(1)$-representations. The possible choices of this lift can be parametrized by an integer $b$, and the resulting matrix factorizations are indeed the $T_b$ in 
(\ref{eq:transition-mf}). Thus, in this case one can arrive at the transition defects in two ways, by pushing down the GLSM identity on the right  into the LG phase, or by lifting the LG-identity defect to the GLSM on the left. The cutoff parameter $N$ in the push-down procedure corresponds to the choice of lift of $\ZZ_d$ to $U(1)$ representations in the lifting procedure. 

Indeed, this does not work in more complicated situations. If 
$P$ appears non-linearly in the superpotential, there is more freedom in the lifting procedure, and not all lifts of the LG-identity defect can be obtained as push-downs of the GLSM identity defect. In this case, only the pushed-down GLSM identity defects yield the correct transition defects.

Note that the choice of $b$ (respectively $N$) precisely determines the $U(1)$-representations appearing in the matrix factorization $T_b$, and hence the representations of GLSM branes obtained by fusing LG branes with $T_b$.
Via fusion, the defects $T_b$ therefore lift the category of D-branes in the LG phase to different subcategories of the category of GLSM branes with restricted $U(1)$-representations. 
Following the terminology  of \cite{HHP_08}, the branes lifted by the defects $T_b$ are automatically grade restricted and the choice of $b$ precisely determines the possible charge windows.

\subsection{Transition defect from the Landau-Ginzburg to the Geometric Phase} \label{sec:LGgeom}

To obtain a transition defect from the LG to the geometric phase, we now have to push down the GLSM on the left side of the defect $T_b$ to the geometric phase. 
Indeed, the push-down of individual D-branes from the GLSM to the geometric phase has been discussed in \cite{HHP_08}. It boils down to the following procedure. Start with a matrix factorization $Q$ of $W(P,X_i)=PG(X_i)$ representing the D-brane in the GLSM. Now regard the $\CC[P,X_i]$-modules $Q_0,Q_1$ as infinite-rank modules over $\CC[X_i]$ and the maps $q_0,q_1$ as maps of $\CC[X_i]$-modules. Then unfold the matrix factorization (regarded as 2-periodic twisted complex) according to $R$-charge into an infinite twisted complex.\footnote{$P$ has $R$-charge $2$.} The twist of this complex is just $G(X_i)$, and dividing out the ideal generated by $G(X_i)$, one arrives at an honest complex of $\CC[X_i]/(G(X_i))$-modules. 
Regarding $\CC[X_i]/(G(X_i))$ as structure sheaf $\mathcal{O}_M$ of the projective hypersurface $M=\{G(X_i)=0\}\subseteq\mathbb{P}^{N-1}$ one obtains an infinite complex of coherent sheaves on $M$, which can be shown to be quasi-isomorphic to a finite one. In this way one obtains an object in the derived category of coherent sheaves on $M$, i.e. a D-brane in the geometric phase from the matrix factorization $Q$.

It was suggested in \cite{HHP_08} that this procedure can be regarded as a variant of Kn\"orrer periodicity. The latter states that the categories of matrix factorizations for a superpotential $W_L(X_i)$ and $W_L(X_i)+uv$ are equivalent. The isomorphism between the two categories can be thought of as tensoring with the matrix factorization
\begin{equation}
	\begin{tikzcd}
		K: \; K_1 \rar[shift left]{u} & \lar[shift left]{v} K_0 ,
	\end{tikzcd}
\end{equation}
which in turn can be written as fusion product with a matrix factorization\footnote{$I$ is the identity matrix factorization.} $I\otimes K$, as 
emphasized in \cite{Carqueville:2012dk}.
Indeed, for $W_L=W(Y_i)$, $u=P$ and $v=-G(X_i)$, fusing $I\otimes K$ with $Q$ indeed sets $G(X_i)=0$ and integrates out $P$ as described above.

In the following we will push down the GLSM on the left of the defect $T_b$ into the geometric phase by applying this Kn\"orrer map. The resulting object, describing a defect from the LG to the geometric phase, will be a hybrid between matrix factorization and coherent sheaf on $M$. It can be used to directly transfer D-branes from the LG to the geometric phase.

The defect $T_b$ is a matrix factorization of 
\begin{equation}\label{eq:wpxy}
	W(P,X_i,Y_i) = P\cdot G(X_i) - G(Y_i)\,.
\end{equation}
We will apply Kn\"orrer periodicity by setting $W_L =-G(Y_i)$, $u=P$ and $v=G(X_i)$. Thus we have to expand the matrix factorization in $P$ and divide by $G(X_i)$.

The expansion takes a very simple form. The modules are expanded according to
\begin{equation}
T_s=T_s^0\oplus P T_s^1\oplus P^2 T_s^2\oplus \ldots\,.
\end{equation}
Note that this is an expansion with respect to $R$-charge.
Since $P$ appears linearly, the maps $t_s$ can be written as
\begin{equation}
t_s=t_s^0+Pt_s^1\,,
\end{equation}
where $t_s^0:T_r^i\rightarrow T_{r+s\,\text{mod }2}^i$ and $t_s^1:T_r^i\rightarrow T_{r+s\,\text{mod }2}^{i+1}$. 
Since $T_b$ is a matrix factorization of (\ref{eq:wpxy}) we have
\begin{equation}
	t_0 \circ t_1 = t_0^0 \circ t_1^0 + P\cdot(t_0^0 \circ t_1^1 + t_0^1 \circ t_1^0) + P^2\cdot t_0^1
	\circ t_1^1 {=}  PG(X_i) - G(Y_i) .
\end{equation} 
Comparing powers of $P$, we obtain the following relations on $t_s^0, t_s^1$
\begin{align}
		t_0^0 \circ t_1^0 &= -G(Y_i) ,  \label{eq:T-knoerrer-relations-mf} \\    
		(t_0^0 \circ t_1^1 + t_0^1 \circ t_1^0) &= G(X_i) ,  \label{eq:T-knoerrer-relations-commutativity}\\
		t_0^1 \circ t_1^1 &= 0 \label{eq:T-knoerrer-relations-complex}.
\end{align}
Next, we divide by the ideal generated by $G(X_i)$. For this, we define
\begin{equation}
	\T_s^\kappa :=T_s^\kappa \big/ ( G(X_i) )  .
\end{equation}
Then, because of (\ref{eq:T-knoerrer-relations-mf}), the maps $t_s^0$ define matrix factorizations
\begin{equation}
	\begin{tikzcd}
		\T_b^\kappa : \; \T_1^\kappa \rar[shift left]{t_1^0} & \lar[shift left]{t_0^0} \T_0^\kappa 
	\end{tikzcd} 
\end{equation} 
of ${W_L=-G(Y_i )}$ over the ring $\CC[X_i,Y_i]/(G(X_i))$
for all $\kappa\in\NN_0$. Moreover, due to 
(\ref{eq:T-knoerrer-relations-commutativity}) and the fact that we divided by $G(X_i)$, 
the diagram
\begin{equation}
	\begin{tikzcd}
		\T_1^\kappa \rar{t_1^0} \dar{t_1^1} & \T_0^\kappa \dar{t_0^0} \\
		\T_0^{\kappa +1} \rar{-t_0^0} & \T_1^{\kappa + 1}
	\end{tikzcd}
\end{equation}
commutes. Hence, the maps $t_s^1$
define an even (i.e.~bosonic) morphism $\widetilde{\varphi}^\kappa:T_b^\kappa\rightarrow \T_b^{\kappa+1}[1]$
of matrix factorizations, or equivalently an odd 
(i.e.~fermionic) morphism $\varphi^\kappa: \T_b^\kappa \rightarrow \T_b^{\kappa+1}$, with 
\begin{equation}
	\varphi^\kappa = \begin{pmatrix}
		0 & t_1^1 \\
		t_0^1 & 0 
	\end{pmatrix} .
\end{equation}
Finally, because of (\ref{eq:T-knoerrer-relations-complex}), the $\varphi^\kappa$ compose to zero, i.e.~$\varphi^{\kappa+1}\circ\varphi^\kappa=0$. Thus we arrive at a semi-infinite complex
\begin{equation}
	\begin{tikzcd}
		\T_b^0 \rar{\varphi^0} & \T_b^1 \rar{\varphi^1} & \T_b^2 \rar{\varphi^2} & \cdots
	\end{tikzcd} 
\end{equation}
of matrix factorization of $-G(Y_i)$ over the ring $\CC[X_i,Y_i]/(G(X_i))$.

Writing this out, one obtains a type of semi-twisted double complex (see for example A.$3$ in
\cite{Eisenbud}), where the rows are the twisted complexes coming from the matrix factorizations
and the vertical maps correspond to the morphisms $\varphi^i$. We have 
\begin{equation}\label{eq:T-double-complex}
	\begin{tikzcd}
		& 0 \dar & 0 \dar & 0 \dar & 0 \dar \\
		\cdots \rar{t_0^0} & \T_1^0 \rar{t_1^0} \dar{t_1^1} & \T_0^0 \rar{t_0^0} \dar{t_0^1} & \T_1^0
		\rar{t_1^0} \dar{t_1^1} & \T_0^0 \dar{t_0^1} \rar{t_0^0} & \cdots \\
		\cdots \rar{-t_1^0} & \T_0^1 \rar{-t_0^0} \dar{t_0^1} & \T_1^1 \rar{-t_1^0} \dar{t_1^1} & \T_0^1
		\rar{-t_0^0} \dar{t_0^1} & \T_1^1 \rar{-t_1^0} \dar{t_1^1} & \cdots \\
		\cdots \rar{t_0^0} & \T_1^2 \rar{t_1^0} \dar{t_1^1} & \T_0^2 \rar{t_0^0} \dar{t_0^1} & \T_1^2
		\rar{t_1^0} \dar{t_1^1} & \T_0^2 \dar{t_0^1} \rar{t_0^0} & \cdots \\
		& \vdots & \vdots & \vdots & \vdots 
	\end{tikzcd}
\end{equation}

The hybrid defect $\T_b$ is then given by the total (twisted) complex of this semi-twisted double complex. The
total complex is a generalization of the cone construction. Its terms are given by the
direct sum over the diagonals of the above double complex and the differentials are
obtained from the horizontal and vertical maps of the double complex.\footnote{The signs in the definition of the differentials of the total complex exactly cancel the signs appearing in the double complex \eqref{eq:T-double-complex}.}

Furthermore, due to the 2-periodicity of the rows in eq.~\eqref{eq:T-double-complex}, the result can be written
as a stack of matrix factorizations of $-G(Y_i)$
\begin{equation}
	\begin{tikzcd}[column sep=large]
		\T_1^0 \ar[shift left]{r}[inner sep=1pt]{t_1^0} \ar{dr}[inner sep=0pt, very near start, swap]{t_1^1}
		\dar[phantom,"\oplus" marking]{} & \ar[shift left]{l}[inner sep=1pt]{t_0^0} \ar{dl}[inner sep =0pt, very near start]{t_0^1} \dar[phantom,"\oplus" marking]{} \T_0^0 \\
		\T_1^1 \ar[shift left]{r}[inner sep=1pt]{t_1^0} \ar{dr}[inner sep=0pt, very near start, swap]{t_1^1}
		\dar[phantom,"\oplus" marking]{} & \ar[shift left]{l}[inner sep=1pt]{t_0^0} \ar{dl}[inner sep =0pt, very near start]{t_0^1} \dar[phantom,"\oplus" marking]{} \T_0^1 \\
		\T_1^2 \ar[shift left]{r}[inner sep=1pt]{t_1^0} \ar{dr}[inner sep=0pt, very near start, swap]{t_1^1}
		\dar[phantom,"\oplus" marking]{} & \ar[shift left]{l}[inner sep=1pt]{t_0^0} \ar{dl}[inner sep =0pt, very near start]{t_0^1} \dar[phantom,"\oplus" marking]{} \T_0^2 \\
		\vdots & \vdots 
	\end{tikzcd}
\end{equation}

For computations it is convenient to write $\T_b$ using the cone construction, i.e.\ we consider the
total complex as a recursive application of the mapping cone. Since $\varphi^\kappa$ induces
a map onto $\cone(\varphi^{\kappa + 1}) $ by mapping to
the first component in $\T_b^{\kappa+1} \oplus \T_b^{\kappa+2}$, we can write $\T_b$ as 
\begin{equation}\label{eq:T-cone}
	\T_b = \cone ( {\varphi }^0: \T_b^0 \rightarrow \cone( {\varphi }^1:
	\T_b^1 \rightarrow \cone ( {\varphi }^2: \T_b^2 \rightarrow \cdots
	))) .
\end{equation}

In contrast to D-branes which become complexes of coherent sheaves when transported from the LG to the geometric phase, $\T_b$ is still a matrix factorization of the
remaining part of the superpotential $W_L = -G(Y_i)$, corresponding to the Landau-Ginzburg
model on the right side of the defect. It is a hybrid object between a matrix factorization and a complex of coherent sheaves.

\subsection{Action on the D-branes}\label{sec:LGgeomaction}

The hybrid defects $\T_b$ constructed in the previous section act via fusion as functors from the category of D-branes in the LG model (equivariant matrix factorizations) into the category of D-branes in the non-linear sigma model (derived category of coherent sheaves on the target space $M$). The action is given by taking the tensor product over the LG model. 

Fusing the defect $\T_b$ with a LG brane, i.e.~taking the tensor product with a matrix factorization of $G(Y_i)$ gives rise to an untwisted complex.\footnote{twists add under the tensor product} Regarding the free $\CC[X_i]/(G(X_i))$-modules in this complex as structure sheaves on $M$, this yields a complex of coherent sheaves, i.e.~a D-brane in the geometric phase described by an object of the derived category of coherent sheaves on $M$.

A convenient way to explicitly compute such a tensor product, is to make use of the fact that the fusion product commutes with the cone construction. So if 
\begin{equation}
	\begin{tikzcd}
		Q: \; Q_1 \rar[shift left]{q_1} & \lar[shift left]{q_0} Q_0 
	\end{tikzcd}
\end{equation}
is a matrix factorization of $G(Y_i)$, then using (\ref{eq:T-cone}), one can write the fusion as 
\begin{equation}\label{eq:TxB-cone}
	\begin{split}
		\T_b * Q &= \cone \left( {\varphi}^0 : \T_b^0 \rightarrow \cone\left( {\varphi}^1 : \T_b^1 \rightarrow \cone\left(\ldots\right) \right) \right) * Q \\
		&= \cone \left(\bar{\varphi}^0 : \T_b^0 * Q\rightarrow \cone\left( \bar{\varphi}^1 : \T_b^1 * Q \rightarrow \cone\left(\ldots\right) \right) \right) \,.
	\end{split}
\end{equation}
Here the maps $\bar{\varphi}^\kappa$ are obtained by pushing down the maps $\varphi^\kappa\otimes\id_{Q}$ to the fusion product by means of  isomorphisms
\begin{equation}
\begin{split}
&r^\kappa_\T:\T_b^\kappa\otimes Q\stackrel{\cong}{\longrightarrow} \T_b^\kappa*Q\,,\quad
(r^*_\T)^\kappa:\T_b^\kappa* Q\stackrel{\cong} {\longrightarrow}\T_b^\kappa\otimes Q\,,\\
&r_\T^\kappa\circ (r^*_\T)^\kappa=\id_{\T_b^\kappa* Q}\,,\quad
(r^*_\T)^\kappa\circ r_\T^\kappa=\id_{\T_b^\kappa\otimes Q}\,,
\end{split}
\end{equation}
i.e.
\begin{equation}
\bar{\varphi}^\kappa=r^\kappa_\T\circ (\varphi^\kappa\otimes\id_{Q})\circ (r^*_\T)^\kappa\,.
\end{equation}
Thus, one can determine $\T_b*Q$ by calculating the fusion products of $\T_b^\kappa*Q$ and forming the successive cones of the maps $\bar{\varphi}^\kappa$.
We will make all this explicit in the example discussed in section~\ref{sec:example}.

\section{Example}\label{sec:example}

Next, we apply the methods presented in the previous sections to a concrete example, the $U(1)$-gauged linear sigma model with chiral superfields $P,X_1,\ldots,X_N$ of $U(1)$-charges $(-d,1\ldots,1)$ and superpotential 
$W=PG(X_1,\ldots,X_N)$. Here, $G$ is a homogeneous polynomial of degree $d$. For concreteness, we choose
\begin{equation}
G(X_1,\ldots,X_N)=\sum_{i=1}^NX_i^d\,,
\end{equation}
but our arguments work in the same way for more general $G$.

\subsection{Lifting the LG phase to the GLSM}
\subsubsection{Construction of the defects $T_b$}

We begin by constructing the defects $T_b$ from the LG phase to the GLSM. This could be done by pushing down the GLSM-identity defect to the LG phase on its right side, as put forward in \cite{BR-new}. Here, we will take an alternative approach and instead lift the LG-identity defect to the GLSM on its left. In the simple case at hand, this yields equivalent results, as pointed out in section \ref{sec:Transition}. 

Defects in the LG phase can be represented by $\ZZ_d\times\ZZ_d$-equivariant matrix factorizations of $G(X_i)-G(Y_i)=\sum_{i=1}^N(X_i^d-Y_i^d)$. The identity defect is represented by a matrix factorization of Koszul type \cite{kapustin_relation_2004}. Its general form is given in (\ref{eq:LG-id}). In this section, we will use a matrix representation of the defect field  $\alpha$ appearing in that formula. More precisely, we introduce the 
$\CC[X_1,\ldots,X_N,Y_1,\ldots,Y_N]$-module
\begin{equation}
\bar{V} = \CC[X_1,\ldots,X_N,Y_1,\ldots,Y_N]\otimes\bigoplus_{\nu = 0}^{d-1} \CC\{[\nu] , [-\nu], 0\}\,,
\end{equation}
whose generators $f_\nu$ of  $\ZZ_d\times \ZZ_d \times U(1)_R$ charge $\{[\nu],[-\nu],0\}$ corrrespond to $\alpha^\nu$ in (\ref{eq:LG-id}). (Here and in the following $[\cdot]$ denotes the rest class modulo $d$.)
Multiplication by $\alpha$ is represented by the shift matrix ${\varepsilon: \bar{V} \rightarrow
\bar{V}}$, 
$\varepsilon(f_{\nu }) = f_{\{\nu+1\}_d}$. 
Here and in the following $\{n\}_d$ denotes the unique representative of the rest class $[n]$ modulo $d$ in $\{0,\ldots,d-1\}$.\footnote{I.e.~$\{n\}_d$ is the unique element in $\{0,\ldots,d-1\}\cap(n+d\ZZ)$.}
The $N$-dimensional $\CC$-vector space $V$ of (\ref{eq:LG-id}) will be denoted $V_I$ here, and its basis vectors $\theta_i$ by $e_i$. They have $\ZZ_d\times\ZZ_d\times U(1)_R$-charge $\{[1],[0],-1+\frac{2}{d}\}$.
With this notation, the 
matrix factorization corresponding to  (\ref{eq:LG-id}) can then be explicitly written as
\begin{equation}\label{eq:id-orb-koszul}
	\begin{tikzcd}
		\orb{I}: \orb{I}_0=\bigwedge^{\text{odd} } V_I \otimes \bar{V} \rar[shift left]{\orb{\imath}_1} 
		& \lar[shift left]{\orb{\imath}_0} \bigwedge^{\text{even}} V_I \otimes \bar{V}=\orb{I}_1
	\end{tikzcd},
\end{equation}
where
\begin{equation}
\orb{\imath}_0=\left(\orb{\delta}_I+\orb{\sigma}_I\right)\Big|_{\bigwedge^{\text{even} } V_I \otimes \bar{V}}\,,\qquad
\orb{\imath}_1=\left(\orb{\delta}_I+\orb{\sigma}_I\right)\Big|_{\bigwedge^{\text{odd} } V_I \otimes \bar{V}}\,,
\end{equation}
with
\begin{equation}
	\begin{split}
		\orb{\delta}_I &= \sum_{i=1}^N \iota_{e_i^*}\otimes(X_i\cdot\II-Y_i\cdot\varepsilon)\\
		\orb{\sigma}_I &= \sum_{i=1}^N  (e_i \wedge \cdot ) \otimes
		\prod_{l=1}^{d-1}( X_i \cdot \II - \xi^l Y_i \cdot \varepsilon).
	\end{split}
\end{equation}
$\xi$ denotes an elementary $d$th root of unity. (One can obtain the differentials from those in  (\ref{eq:continuous-id}) by plugging in the explicit superpotential and computing the quotients of differences.)

Lifting this matrix factorization to the GLSM on the left
\begin{equation}
\begin{tikzpicture}
	\node[LG] (I-LHS) {$\sum_i X_i^d$};
	\node[LG,anchor=west] (I-RHS) at (I-LHS.east) {$\sum_i Y_i^d$};
	\draw[blue!20] (I-LHS.north east) -- (I-LHS.south east);
	\draw[dashed] (I-LHS.north east) -- (I-LHS.south east);
	\node[anchor=north] (I) at (I-LHS.south east) {$\orb{I}$};

	\node[GLSM] (T-LHS) [right=of I-RHS] {$P\sum_i X_i^d$};
	\node[LG,anchor=west] (T-RHS) at (T-LHS.east) {$\sum_i Y_i^d$};
	\draw[blue,very thick] (T-LHS.north east) -- (T-LHS.south east);
	\draw[->,decorate,decoration={snake,amplitude=.4mm,segment length=2mm,post length=1mm}] (I-RHS) --
	(T-LHS);
	\node[anchor=north] (T) at (T-LHS.south east) {$T_b$};
\end{tikzpicture}
\end{equation}
means constructing a $U(1) \times \ZZ_d \times U(1)_R$-equivariant
matrix factorization of the superpotential $PG(X_i) - G(Y_i)$ which reduces to the factorization
$\orb{I}$ when taking the Landau-Ginzburg limit. 
On the level of matrix factorizations, taking this limit
comprises shifting the $U(1)_R$-charges according to
\begin{equation}
q_R\mapsto q_R+\frac{2}{d}q_{U(1)}\,,
\end{equation} 
setting $P=1$, and breaking the $U(1)$-symmetry to $\ZZ_d$.
Thus, 
lifting the matrix factorization $\orb{I}$ can be achieved by inserting factors of $P$ 
at appropriate places in the matrices $\orb{\imath}_i$
so as to obtain a factorization of $PG(X_i)-G(Y_i)$, lifting the $\ZZ_d$-representation to a $U(1)$-representation and shifting the R-charges according  to 
\begin{equation}
q_R\mapsto q_R-\frac{2}{d}q_{U(1)}\,.
\end{equation} 
Indeed, there is one such lift for every $b\in\ZZ$:
\begin{equation}\label{eq:T-koszul}
	\begin{tikzcd}
		T_b: \displaystyle\bigoplus_{k \text{ odd}} \textstyle\bigwedge^{k } V_T \otimes \glsm{V}^{b,k} \rar[shift left]{t_1} 
		& \lar[shift left]{t_0} \displaystyle\bigoplus_{k \text{ even}} \textstyle\bigwedge^{k } V_T \otimes \glsm{V}^{b,k}	
	\end{tikzcd}.
\end{equation}
Here, $V_T$ is an $N$-dimensional $\CC$ vector space, which lifts $V_I$. 
By abuse of notation we denote its basis vectors by the same symbols $e_1,\ldots,e_N$ as the ones of $V_I$. Moreover, $\glsm{V}^{b,k}$ are rank-$d$ free $\CC[X_1,\ldots,X_N,Y_1,\ldots,Y_N,P]$-modules lifting the $\bar{
V}$. We will denote their generators by 
 $f_\mu^{b,k}$, $\mu\in\{0,\ldots,d-1\}$.
 
The maps $t_s$ are obtained by replacing 
\begin{equation}
X_i\cdot\II \longmapsto X_i\cdot I_P^{[k]}
\end{equation}
 in the formulas (\ref{eq:id-orb-koszul}),
 where 
 \begin{equation}
 I_P^{[k]}=
	\begin{pmatrix}
		1 &        &   &   &   &       &   & \\
		  & \ddots &   &   &   &       &   & \\
		  &        & 1 &   &   &       &   & \\
			&        &   & P &   &       &   & \\
			&        &   &   & 1 &       &   & \\
			&        &   &   &   &\ddots &   & \\
			&        &   &   &   &       & 1 &  
	\end{pmatrix}
 \end{equation}
is the identity matrix whose $(d-k)$th diagonal entry is replaced by $P$. Note here that
$\varepsilon^{-1}\cdot I_P^{[k]}\cdot\varepsilon=I_P^{[k+1]}$ and that
\begin{equation}
\prod_{l=0}^{d-1}(X\cdot I_P^{[k+l]}-\xi^lY\cdot \varepsilon)=\left(P\,X^d-Y^d\right)\II \,.
\end{equation} 
Concretely, 
\begin{equation}\label{eq:ts}
	t_s = (\delta_T + \sigma_T )\Big|_{\bigoplus_{k + s \text{ even}} \bigwedge^k V_T \otimes
	\glsm{V}^{b,k}}\,,
\end{equation}
where $\delta_T$ and $\sigma_T$ act on $\textstyle\bigwedge^{k } V_T \otimes \glsm{V}^{b,k}$
as
\begin{align}\label{eq:T-delta-sigma}
	\delta_T\Big|_{\textstyle\bigwedge^{k } V_T \otimes \glsm{V}^{b,k}}=:
	\delta_T^k &= \sum_i \iota_{e_i^*} \otimes \left(X_i\,I_P^{[k-1]} - Y_i\,\varepsilon\right) \\
	\sigma_T\Big|_{\textstyle\bigwedge^{k } V_T \otimes \glsm{V}^{b,k}}=:\sigma_T^k &= \sum_{j=1}^N (e_j \wedge \cdot ) \otimes \left( \prod_{l=1}^{d-1 } ( X_j \cdot 
	I_P^{[k+l]} - \xi^l Y_j \cdot \varepsilon) \right) .
\end{align}
The $\ZZ_d \times \ZZ_d \times U(1)_R$-representation on $\orb{I}$ is lifted to a $U(1)\times
\ZZ_d \times U(1)_R$-representation on $T_b$. The corresponding charges of $e_i$ are given by 
$\{1,0,-1\}$, and the  $f_\mu^{b,k}$ carry charges
\begin{equation}
\Big\{ 
			b + \mu - d \cdot \left\lfloor \frac{\mu + k}{d} \right\rfloor , 
			[-b -\mu], 
			 -\frac{2}{d}( b + \mu ) + 2 \cdot \left\lfloor \frac{\mu + k}{d} \right\rfloor  \Big\}\,.
\end{equation}
Then, the generators $e_{i_1}\wedge\ldots\wedge e_{i_k} \otimes f_\mu^{b,k}$ of
$\bigwedge^k V_T \otimes \glsm{V}^{b,k}$ have charges 
\begin{eqnarray}
 &&\Big\{ 
			b + \mu + k - d \cdot \left\lfloor \frac{\mu + k}{d} \right\rfloor , 
			[-b -\mu], 
			 -\frac{2}{d}( b + \mu ) + 2 \cdot \left\lfloor \frac{\mu + k}{d} \right\rfloor - k \Big\} \\
&&\qquad\qquad		= \Big\{ 
			b + \{\mu + k\}_d , 
			[-b -\mu], 
			 -\frac{2}{d}( b + \mu ) + 2 \cdot \left\lfloor \frac{\mu + k}{d} \right\rfloor - k \Big\} \nonumber
\end{eqnarray}
Altogether, the matrix factorization (\ref{eq:T-koszul}) is a concrete representation of  (\ref{eq:transition-mf}) in the example at hand. 

Note, that the $U(1)$ charges appearing in 
$T_b$ all lie within the set 
$$
\N_{b+d-1}=\{b,b+1,\ldots,b+d-1\}
$$ 
of $d$ consecutive integers with minimum $b$. 
This is called a charge window in \cite{HHP_08}. Indeed, in \cite{HHP_08,hori_exact_2013} D-brane transport between different phases of a GLSM was analyzed using different techniques. By a careful analysis of the gauge sector of the GLSM, and by analyzing the convergence properties of partition functions, it was proposed in \cite{HHP_08,hori_exact_2013} that a smooth transport of D-branes from one phase to another involves a lift of D-branes to the GLSM, whose charges are restricted to such charge windows.  The choice of window corresponds to a choice of path between the phases, as explained in \cite{HHP_08,hori_exact_2013,RlFC_brane-transport_18,Knapp:2016rec}.

Since the $U(1)$-charges of  the defects $T_b$ are contained in a charge window, its fusion with all LG branes are automatically ``grade restricted'' in the sense of \cite{HHP_08,hori_exact_2013}. 
The defect $T_b$ therefore acts as a functor from the category of D-branes of the LG phase to a grade-restricted subcategory of the category of GLSM branes, which is determined by $b$. Even though our analysis does not involve the gauge sector, it still recovers the important property of grade restriction obtained in \cite{HHP_08}.

\subsubsection{Lifting LG-branes by fusion with $T_b$}\label{sec:GLSM-action-on-D-branes}

Next, we exemplify how to lift D-branes from the Landau-Ginzburg phase to the GLSM by fusing them with the defects $T_b$. For this we choose Landau-Ginzburg branes  which are represented by tensor products of linear matrix factorizations. These are Koszul-type matrix factorizations, which can be described as follows.
Let
$V_B$ be an $N$-dimensional vector space with basis vectors $g_i$, $i=1,\ldots,N$ of $\ZZ_d\times U(1)_R$-charges $\{[1],-1+\frac{2}{d}\}$. Furthermore, let $S_{(Y)}=\CC[Y_1,\ldots,Y_N]$ and $S_{Y}\{[c],\frac{2c}{d}\}$ the free rank-$1$ $S_{(Y)}$-module whose $\ZZ_d\times U(1)_R$ charges are shifted by $\{[c],\frac{2c}{d}\}$. Then for each $c\in\ZZ$ there is a matrix factorization
\begin{equation}\label{eq:B-koszul}
	\begin{tikzcd}
		B_{c}: \bigwedge^{\text{odd} } V_B\otimes S_{(Y)}\{[c],\frac{2c}{d}\} \rar[shift left]{\delta_B + \sigma_B} 
		& \lar[shift left]{\delta_B + \sigma_B} \bigwedge^{\text{even}} V_B\otimes S_{(Y)}\{[c],\frac{2c}{d}\} 
	\end{tikzcd}
\end{equation}
with 
\begin{equation}
	\delta_B = \sum_{i=1}^N\iota_{g_i^*}\otimes Y_i, \quad \sigma_B = \sum_{i=1}^N ( g_i \wedge
	\cdot )\otimes Y_i^{d-1}.
\end{equation}

The fusion product of $T_b$ and $B_c$ is given by the $\ZZ_d$-invariant part of the tensor product of the respective matrix factorizations:
\begin{equation}
	T_b * B_c = \left( T_b \otimes B_c\right)^{\ZZ_d},
\end{equation}
Here, the tensor product $T_b \otimes B_{c}$ is given by 
\begin{equation}\label{eq:TxB-koszul}
	\begin{tikzcd}[column sep=2em, scale cd=0.8]
		\displaystyle\bigoplus_{k+l \text{ odd}} \left( \textstyle\bigwedge^{k } V_T \otimes \glsm{V}^{b,k} \otimes \bigwedge^l V_B \otimes S_{(Y)}\{[c],\frac{2c}{d}\}\right) \rar[shift left]{} 
		& \lar[shift left]{} \displaystyle\bigoplus_{k+l \text{ even}} \left( \textstyle\bigwedge^{k } V_T \otimes \glsm{V}^{b,k} \otimes \bigwedge^l V_B	\otimes S_{(Y)}\{[c],\frac{2c}{d}\}\right).
	\end{tikzcd}
\end{equation}
Note that this tensor product is taken over the polynomial ring $S_{(Y)}=S[Y_1,\ldots,Y_N]$ of fields of the intermediate model.
Generators $e_{i_1}\wedge\ldots\wedge e_{i_k} \otimes f^{b,k}_\nu \otimes g_{j_1}\wedge\ldots\wedge g_{j_l}$ of $\bigwedge^{k } V_T \otimes \glsm{V}^{b,k}_\nu
\otimes \bigwedge^l V_B\otimes S_{(Y)}\{[c],\frac{2c}{d}\}$ have $U(1)\times\ZZ_d\times U(1)_R$-charges 
\begin{equation}
	\bigg\{ b + k + \nu - d \left\lfloor \frac{\nu + k}{d}
	\right\rfloor ,\; [-b -\nu + c + l],\; 
\frac{2}{d}( c - b - \nu ) + 2 \cdot \left\lfloor \frac{\nu + k}{d} \right\rfloor - k -
	l\bigg\}. 
\end{equation}

A convenient way to calculate a finite rank matrix factorization isomorphic to this tensor product is via associated Cohen-Macaulay modules (c.f.~\ref{sec:Defects_GLSM_LG}, for more details see \cite{Roggenkamp-permutation-branes,BR-LG-defects}).

The matrix factorization $T_b$ can be obtained from the $U(1)
\times \ZZ_d \times U(1)_R$-equivariant $\CC[P,X_i,Y_i]/(P\cdot G(X_i)-G(Y_i)
)$-free resolution of the $\CC[P,X_i,Y_i]/( P\cdot G(X_i)-G(Y_i))$-module 
\begin{equation}
	M_T = \coker(t_1) \cong \coker\left( (\delta_T + \sigma_T)\Big|_{\bigoplus_{k \text{ odd}} \bigwedge^k V_T
	\otimes \glsm{V}^{b,k}\otimes \CC[P,X_i,Y_i]/(P\cdot G(X_i)-G(Y_i))} \right).
\end{equation}
The matrix factorization 
$B_c$ can be obtained from the $\ZZ_d$-equivariant $\CC[Y_i]/( G(Y_i))$-free resolution of the $\CC[Y_i]/( G(Y_i)
)$-module
\begin{equation}
	M_B = \coker \left(\delta_B\Big|_{\bigwedge^1 V_B\otimes S_{(Y)}\{[c],\frac{2c}{d}\}}\right) \cong \CC[Y_1,\ldots,Y_N]\{[c],\textstyle \frac{2c}{d}\}\big/ (
	Y_1,\ldots,Y_N )\,.
\end{equation}

A matrix factorization isomorphic to the tensor product $T_b \otimes B_c$ can now be obtained from the tensor product $M_T\otimes_{S_{(Y)}}M_B$. The latter is a $ \CC[P,X_i]/( P\cdot G(X_i))$-module, which is isomorphic to 
\begin{equation}
	M_T \otimes_{\CC[Y_1,\ldots,Y_N]} M_B \cong  \coker\left(
	t_1\Big|_{Y_1=\ldots=Y_N=0} \right)\,.
\end{equation}
From this it is not difficult to read off a finite-rank matrix factorization isomorphic to $T_b\otimes B_c$, which we denote by $(T_b\otimes B_c)^\text{red}$:
\begin{equation}\label{eq:TxBred}
	\begin{tikzcd}[column sep=large]
		{\displaystyle\bigoplus_{k\text{ odd}}} \bigwedge^{k}
		V_{T}\otimes \widetilde{V}^{b,k}\{0,[c],\frac{2c}{d}\} \rar[shift
		left]{\delta_{\text{red}} + \sigma_{\text{red}}} 
		& \lar[shift left]{\delta_{\text{red}} + \sigma_{\text{red}}} \displaystyle\bigoplus_{k\text{
		even}} \textstyle\bigwedge^{k} V_{T}\otimes \widetilde{V}^{b,k}\{0,[c],\frac{2c}{d}\}
	\end{tikzcd}
\end{equation}
with
\begin{align}\label{eq:TxB-red-delta-sigma}
	\delta_{\red}^k &:=\delta_{\red}\Big|_{\bigwedge^{k}
		V_{T}\otimes \widetilde{V}^{b,k}\{0,[c],\frac{2c}{d}\}}
	= \sum_{i=1}^N\iota_{e_i^*} \otimes X_i\,I_P^{[k-1]} \\
	\sigma_\red^k &:=\sigma_{\red}\Big|_{\bigwedge^{k}
		V_{T}\otimes \widetilde{V}^{b,k}\{0,[c],\frac{2c}{d}\}}
	= \sum_{j=1}^N (e_j \wedge \cdot ) \otimes \left( X_j^{d-1} \prod_{a=1}^{d-1 } I_P^{[k+a]} \right)
\end{align}
and $\widetilde{V}^{b,k}=\glsm{V}^{b,k}\otimes_{S_{(Y)}}\CC[P,X_1,\ldots,X_N,Y_1,\ldots,Y_N]/( Y_1,\ldots,Y_N)$.

The generators $e_{i_1}\wedge\ldots\wedge e_{i_k} \otimes \widetilde{f}_\mu^{b,k}$ of $\bigwedge^{k}
		V_{T}\otimes \widetilde{V}^{b,k}\{0,[c],\frac{2c}{d}\}$ have charges 
\begin{equation}\label{eq:gencharges}
 \bigg\{ b + k + \mu - d \left\lfloor \frac{\mu + k}{d} \right\rfloor , [-b -\mu + c] ,
 \frac{2}{d}( c - b - \mu ) + 2 \left\lfloor \frac{\mu + k}{d} \right\rfloor - k\bigg\}. 
\end{equation}
Invariance under the squeezed-in $\ZZ_d$ gauge group requires 
\begin{equation}
	[-b - \mu + c] = 0 \quad \Rightarrow \quad \mu = \{c-b\}_d\,,
\end{equation}
which means that the only generators surviving the projection onto the $\ZZ_d$-invariant part are
$\bar{e}_{(i_1,\ldots,i_k)}:=e_{i_1}\wedge\ldots\wedge e_{i_k} \otimes \widetilde{f}_{\{c-b\}_d}^{b,k}$. 
The fusion $T_b*B_c$ is then isomorphic to the finite rank matrix factorization $(T_b*B_c)^\text{red}$ defined by
\begin{equation}
	\begin{tikzcd}
		\bigoplus_{k\,\text{odd}}\bigwedge^{k} V_{(T\otimes B)} \otimes S_{(X,P)}\{q^k,q^k_R\}\rar[shift
		left]{\delta + \sigma} 
		& \lar[shift left]{\delta + \sigma}  \textstyle\bigoplus_{k\,\text{odd}}\bigwedge^{k} V_{(T\otimes B)}\otimes S_{(X,P)}\{q^k,q^k_R\}\,.
	\end{tikzcd}
\end{equation}
Here $V_{(T\otimes B)}$ is an $N$-dimensional vector space with basis vectors $\widetilde{e}_i$ of $U(1)\times U(1)_R$-charges $\{1,-1\}$, 
\begin{equation}\label{eq:qqr}
(q^k,q^k_R)=\scriptstyle\left(b  + \{c-b\}_d - d \left\lfloor \frac{\{c-b\}_d + k}{d} \right\rfloor , 
 \frac{2}{d}( c - b - \{c-b\}_d ) + 2 \left\lfloor \frac{\{c-b\}_d + k}{d} \right\rfloor \right)\,,
\end{equation}
$S_{(X,P)}=\CC[X_1,\ldots,X_N,P]$, and 
$\delta$ and $\sigma$ are given by 
\begin{equation}
	\begin{split}
		\delta^k &:=\delta\Big|_{\bigwedge^{k} V_{(T\otimes B)} \otimes S_{(X,P)}\{q^k,q^k_R\}}
		= \begin{cases}
			\sum_{i=1}^N
			\iota_{\widetilde{e}_i^*}\otimes X_i &\quad \text{for $[k-1] \ne d-1-[c-b]$} \\
			\sum_{i=1}^N\iota_{\widetilde{e}_i^*}\otimes PX_i &\quad \text{for $[k-1] = d-1-[c-b]$} \\
		\end{cases}\\
		\sigma^k &:=\sigma\Big|_{\bigwedge^{k} V_{(T\otimes B)} \otimes S_{(X,P)}\{q^k,q^k_R\}}
		= \begin{cases}
			\sum_{i=1}^N  (\widetilde{e}_i\wedge\cdot)\otimes PX_i^{d-1}&\quad \text{for $[k] \ne d-1-[c-b]$} \\
			\sum_{i=1}^N  (\widetilde{e}_j\wedge\cdot)\otimes X_i^{d-1}&\quad \text{for $[k] = d-1-[c-b]$} 
		\end{cases}
	\end{split}
\end{equation}

\subsection{Transition defect to the geometric phase}

Next, we will compute an explicit form of the hybrid defects $\T_b$ describing the transition from LG to the geometric phase. Starting point are the defects $T_b$ lifting the LG phase into the GLSM, which have been discussed in the previous subsection. They have a Koszul-type representation given in (\ref{eq:T-koszul}) with 
maps (\ref{eq:ts}) written in terms of $\delta_T$ and $\sigma_T$. The latter can be decomposed into the components
\begin{eqnarray}
	\delta_T^k &:=&\delta_T\Big|_{\bigwedge^kV_T\otimes \glsm{V}^{b,k}}= \sum_{i=1}^N \iota_{e_i^*} \otimes \left(
	X_i I_P^{[k-1]} -  Y_i \varepsilon\right) \\
	\sigma_T^k &:=&\sigma_T\Big|_{\bigwedge^kV_T\otimes \glsm{V}^{b,k}}= \sum_{i=1}^N (e_i \wedge \cdot ) \otimes \left( \prod_{a=1}^{d-1 } ( X_i \; I_P^{[k+a]} - \xi^a Y_i \; \varepsilon) \right).
\end{eqnarray}
As outlined in section~\ref{sec:Transition}, we separate $\delta_T$ and $\sigma_T$ into maps of $P$-degree $0$ and $1$, which
we denote by $\delta_\T$ and $\sigma_\T$, respectively:
\begin{equation}\label{eq:delta_T-separation}
	\begin{split}
		\delta_T &= (\delta_T)^0 + P \cdot (\delta_T)^1 =: \delta_\T + P \cdot \rho_\T \\
		\sigma_T &= (\sigma_T)^0 + P \cdot (\sigma_T)^1 =: \sigma_\T + P \cdot \vartheta_\T.
	\end{split}
\end{equation}
The degree $0$ maps $\delta_\T$ and $\sigma_\T$ define matrix factorizations on the modules
\begin{equation}
	\T_s^\kappa = \displaystyle\bigoplus_{k + s \text{ even}} \textstyle\bigwedge^{k }
	V_T \otimes
		\left( \glsm{V}^{b,k}\big/ ( P, G(X_i))\glsm{V}^{b,k} \right) \{-d\kappa,[0],2\kappa\}
\end{equation}
whose components are given by 
\begin{align}\label{eq:frakT-delta-sigma}
	\delta_\T^k &= \sum_{i=1}^N \iota_{e_i^*} \otimes \left(
	X_i I_P^{[k-1]}\Big|_{P=0} -  Y_i \varepsilon\right)\\
	\sigma_\T^k	&= \sum_{i=1}^N (e_i \wedge \cdot ) \otimes \left( \prod_{a=1}^{d-1 } ( X_i \; \left(I_P^{[k+a]}\Big|_{P=0}\right) - \xi^a Y_i \; \varepsilon) \right).
\end{align}
The degree $k$ generators $e_{i_1}\wedge\ldots\wedge e_{i_k} \otimes f_\mu^{b,k}$ of $\T_b^\kappa$ now have $U(1)\times\ZZ_d\times U(1)_R$-charges 
\begin{equation}
		 \bigg\{ b + k + \mu - d \left( \kappa + \cdot \left\lfloor
		\frac{\mu + k}{d} \right\rfloor \right) , [-b -\mu] ,
		 -\frac{2}{d}(b + \mu) + 2 \left( \kappa +
		\left\lfloor \frac{\mu + k}{d} \right\rfloor \right) - k \bigg\}. 
\end{equation}
The maps $\rho_\T$ and $\vartheta_\T$ define the fermionic morphism 
\begin{equation}\label{eq:fermmor}
	\varphi^\kappa = \rho_\T + \vartheta_\T : \; \T_b^\kappa \rightarrow \T_b^{\kappa + 1}\,.
\end{equation}
For instance, 
\begin{equation}
	\rho_\T^k = (\delta_T^k - \delta_\T^k)/P = \sum_{i=1}^N\iota_{e_i^*} \otimes X_i\left(\begin{smallmatrix}
		0 &        &   &        &   \\
		  & \ddots &   &        &   \\
			&        & 1 &        &   \\
			&        &   & \ddots &   \\
			&        &   &        & 0
	\end{smallmatrix}\right)\,.
\end{equation}
The $1$ in the matrix in the last equation is at the 
 $d-\{k-1\}_d$-th position. $\vartheta_\T^k$ is obtained in an analogous way. 
\\

The overall defect $\T_b$ is then built out of the factorizations $\T_b^\kappa$ and morphisms
$\varphi^\kappa$ by a recursive cone construction, as in equation (\ref{eq:T-cone}). By construction, it is
a matrix factorization of infinite rank. In principle, it can be reduced to finite rank 
\cite{Knoerrer,HHP_08}, but we will not do it here. Indeed, we find it more convenient to work with the infinite rank representation, particularly when calculating the fusion with Landau-Ginzburg branes, which will be the focus of the next section.

\subsection{D-Brane Transport  to the geometric phase}

The defects $\T_b$ describe the transition from the Landau-Ginzburg into the geometric phase. 
In particular, the behavior of a D-brane in the LG phase under this transition is described by fusion with $\T_b$. 
$\T_b$ is a hybrid between a matrix factorization of $-G(Y_i)$ and a complex of coherent sheaves on the hypersurface $M = \{G(X_i)=0\} \subset \mathbb{P}^{N-1}$. Its fusion with a matrix factorization of $G(Y_i)$ representing a LG brane, results in a complex of coherent sheaves representing a D-brane in the geometric phase.
\\

Here, we will explicitly compute the fusion of $\T_b$ with the Landau-Ginzburg branes $B_{c}$ introduced in
 ~\eqref{eq:B-koszul},
following the strategy outlined in section \ref{sec:LGgeomaction}. More precisely, we will compute the fusion using formula (\ref{eq:TxB-cone}) for the case $Q=B_c$. As a first step we have to calculate the fusion $\T_b^\kappa*Q$
of the matrix factorizations $\T_b^\kappa$ with $Q$. The fusion product is nothing but the $\ZZ_d$-invariant part of the tensor product (taken over the ring of variables corresponding to fields in the intermediate model)
\begin{equation*}
	\T_b^\kappa * B_{c} = \left( \T_b^\kappa \otimes B_{c} \right)^{\ZZ_d }. 
\end{equation*}
As mentioned before, this kind of tensor product usually yields matrix factorizations of infinite rank, which can however be reduced to isomorphic finite rank matrix factorizations $(\T_b^\kappa * B_{c})^\text{red}$.

We would like to compute the fusion $\T_b* Q$ using the finite rank representations $(\T_b^\kappa * B_{c})^\text{red}$ of $\T_b^\kappa * B_{c}$. In order to perform the successive cone construction of formula (\ref{eq:TxB-cone}) using the reduced representatives, we need  
explicit isomorphisms 
\begin{equation}
	\begin{tikzcd}
		T_b^{} \otimes B_{c} \rar[shift left]{r} 
		& \lar[shift left]{r^*} (T_b^{} \otimes B_{c})^\red
	\end{tikzcd}
\end{equation}
between the unreduced and reduced tensor products discussed in section~\ref{sec:GLSM-action-on-D-branes}. 
These can be obtained by lifting the isomorphisms between Cohen-Macaulay modules associated to the matrix factorizations $T_b\otimes B_c$ and $(T_b\otimes B_c)^{\text{red}}$ to their free resolutions. Indeed, the isomorphisms between the modules associated to unreduced and reduced tensor products was explicitly given in the construction of $(T_b\otimes B_c)^{\text{red}}$ in
section~\ref{sec:GLSM-action-on-D-branes}. Lifting them gives rise to the isomorphisms $r$ and $r^*$ of matrix factorizations below. We refer to appendix~\ref{sec:app} for a derivation. 

Recall that the tensor product matrix factorization $T_b\otimes B_c$
is built on the module
\begin{equation}
\bigoplus_{k,l}\textstyle\bigwedge^kV_T\otimes \widehat{V}^{b,k}\otimes \textstyle\bigwedge^l V_B\otimes S_{(Y)}\{0,[c],\frac{2c}{d}\}\,,
\end{equation}
c.f.~equation~(\ref{eq:TxB-koszul}), whereas the reduced one $(T_b\otimes B_c)^\text{red}$ is built on 
\begin{equation}
\bigoplus_k\textstyle\bigwedge^{k}V_T\otimes\widetilde{V}^{b,k}\{0,[c],\frac{2c}{d}\}\,,
\end{equation}
c.f.~equation~(\ref{eq:TxBred}).
Note that we have not yet projected onto the $\ZZ_d$-invariant part, i.e.~we have not performed the orbifold in the intermediate model yet. These modules have generators\footnote{as modules over 
$\CC[X_1,\ldots,X_N,P]$}
\begin{equation}
e_I\otimes Y^M\,f_\mu^{b,k}\otimes g_J=(e_{i_1}\wedge\ldots\wedge e_{i_k})\otimes 
Y_1^{m_1}\ldots Y_N^{m_N}f_\mu^{b,k}\otimes g_{j_1}\wedge\ldots\wedge g_{j_l}
\end{equation}
in case of the unreduced ones and
\begin{equation}
e_I\otimes \widetilde{f}_\mu^{b,k}=(e_{i_1}\wedge\ldots\wedge e_{i_k})\otimes\widetilde{f}_\mu^{b,k}
\end{equation}
for the reduced ones. 

On the level of the generators, the isomorphism $r$  acts as
\begin{equation}\label{eq:defr}
r(e_I\otimes Y^M\,f_\mu^{b,k}\otimes g_J)=\begin{cases}
		0&\quad\text{if }|J|>0\,\lor\,|M|>0\\
		{e}_I \otimes \widetilde{f}_\mu^{b,k} &\quad \text{else}
	\end{cases}
\end{equation}
and $r^*$ is defined by 
\begin{equation}\label{eq:defr*}
\begin{split}
		r^* ({e}_I \otimes \widetilde{f}_\mu^{b,k}) & = 		
		\Bigg[ (e_{i_1}\otimes\II+g_{i_1}\otimes\varepsilon)\wedge\ldots\wedge (e_{i_k}\otimes\II +g_{i_k}\otimes\varepsilon)
	\wedge
		\left( \sum_{n=0}^N \frac{(\omega^k)^{\wedge n}}{n!}
		\right) \Bigg] (1\otimes f_\mu^{b,k}),
\end{split}
\end{equation}
where $\omega^k$ is given by
\begin{equation}
	\omega^k = \sum_{j=1}^N e_j\wedge g_j \otimes \Lambda_j^k
\end{equation}
with
\begin{equation}
	\Lambda_j^k = \frac{1}{Y_j }\left( \prod_{a=1}^{d-1} (X_j \cdot I_P^{[k+a]} - \xi^a Y_j \cdot \varepsilon) - X_j^{d-1} \left( \prod_{a=1}^{d-1} I_P^{[k+a]} \right) \right).
\end{equation}
Here we treated the $g_i$ and $e_j$ as Grassmann variables, i.e. we used the notation 
\begin{equation}
(g_i\otimes a)(e_j\otimes b)=-e_j\otimes (ab)\otimes g_j\,.
\end{equation}

\subsubsection{Transition Defect $LG \rightarrow geometric$}

We are now ready to discuss the desired transition of D-branes from the LG to the geometric phase, i.e.~to calculate the fusion $\T_b*B_c$. The strategy is to determine the reduced matrix factorizations $(\T_b^{\kappa} * B_{c})^{\text{red}}$ and then, employing the isomorphisms $r$ and $r^*$ presented above, assemble them into the successive cone as in equation (\ref{eq:TxB-cone}).

Since the maps $r$ and $r^*$ are homogeneous in $P$ of degree $0$, they can be pushed down to maps\footnote{Since these maps do not explicitly depend on $\kappa$, we suppress their $\kappa$-dependence.}
\begin{equation}
	\begin{tikzcd}
		\T_b^{\kappa} \otimes B_{c} \rar[shift left]{r_\T} 
		& \lar[shift left]{r_\T^*} (\T_b^{\kappa} \otimes B_{c})^\red
	\end{tikzcd}\,,
\end{equation}
where the finite rank factorization $(\T_b^{\kappa} \otimes
B_{c})^\red$ are given by
\begin{equation}\label{eq:frakTxB-red}
	\begin{tikzcd}[column sep=large]
		(\T_b^\kappa \otimes B_{c})^{\mathrm{red}}: {\displaystyle\bigoplus_{k\text{ odd}}}
		\bigwedge^{k} V_{T}\otimes \widetilde{V}^{b,k,\kappa}\{0,[c],\frac{2c}{d}\} \rar[shift
		left]{\delta_{\text{red}} + \sigma_{\text{red}}} 
		& \lar[shift left]{\delta_{\text{red}} + \sigma_{\text{red}}} \displaystyle\bigoplus_{k\text{
		even}} \textstyle\bigwedge^{k} V_{T}\otimes \widetilde{V}^{b,k,\kappa}\{0,[c],\frac{2c}{d}\}
	\end{tikzcd}\,.
\end{equation}
Here 
\begin{equation}
\widetilde{V}^{b,k,\kappa}=\widetilde{V}^{b,k}/(P,G(X_i))\widetilde{V}^{b,k}\{-d\kappa,[0],2\kappa\}\,,
\end{equation}
and the maps $\delta_\red$ and $\sigma_\red$ are defined by 
\begin{align}\label{eq:cTxB-red-delta-sigma}
	\delta_{\red}\Big|_{\bigwedge^{k} V_{T}\otimes \widetilde{V}^{b,k,\kappa}}&=:\delta_{\red}^{k} = \sum_{i=1}^N\iota_{e_i^*} \otimes X_i I_P^{[k-1]}\Big|_{P=0} 
	= \iota_X \otimes \left(\begin{smallmatrix}
		1 & & & & \\
		 &\ddots& & & & &\\
		  & & 1 & & & & \\
		  & & & 0 & & & \\
		  & & & & 1 & & \\
			& & & & &\ddots & \\
			& & & & & & 1 
	\end{smallmatrix} \right)\\
	\sigma_{\red}\Big|_{\bigwedge^{k} V_{T}\otimes \widetilde{V}^{b,k,\kappa}}&=:\sigma_\red^k = \sum_{j=1}^N (e_j \wedge \cdot ) \otimes \left( X_j^{d-1} \prod_{a=1}^{d-1 } I_P^{[k+a]}\Big|_{P=0} \right) \\
	&= \sum_{j=1}^N (e_j \wedge \cdot ) \otimes
	X_j^{d-1} \left(\begin{smallmatrix}
		0 & & & & \\
		 &\ddots& & & & &\\
		  & & 0 & & & & \\
		  & & & 1 & & & \\
		  & & & & 0 & & \\
			& & & & &\ddots & \\
			& & & & & & 0 
	\end{smallmatrix} \right).
\end{align}
The diagonal entries $0$ respectively $1$ in the matrices in the two equations above are at position $d-k-1$ and $d-k$, respectively.

\medskip
Next, we determine the actual fusion product $(\T_b^\kappa *B_c)^\text{red}$ from the tensor product, by projecting onto the part invariant under the 
gauge group $\ZZ_d$ of the intermediate LG model.

The $U(1)
\times \ZZ_d \times U(1)_R$-charges 
of the generators $e_{i_1}\wedge\ldots\wedge e_{i_k} \otimes \widetilde{f}_\mu^{b,k}$ of $(T_b\otimes B_c)^\red$ have been given in (\ref{eq:gencharges}).
One easily reads off that exactly the generators 
with $\mu=[c-b]$ are $\ZZ_d$-invariant, and hence generate 
the modules of $(T_b*B_c)^\text{red}$. We denote them by $\bar{e}_I={e}_I\otimes \widetilde{f}^{b,k}_{[c-b]}$, and they have $U(1)\times U(1)_R$-charges 
\begin{equation}\label{eq:frakT*B-charges}
\scriptstyle
	\begin{split}
		 &\scriptstyle \bigg\{ b + \{c-b\}_d + k - d \left( \kappa + \left\lfloor \frac{\{c-b\}_d +
		k}{d}\right\rfloor  \right),  
		\frac{2}{d}(c-b - \{c-b\}_d ) + 2\left( \kappa + \left\lfloor \frac{\{c-b\}_d +
		k}{d}\right\rfloor  \right) \bigg\} \\
		&\qquad\qquad\qquad= \left\{ b + \{c-b + k\}_d - d\cdot \kappa , 2 \left( \kappa + \left\lfloor \frac{c-b + k}{d}\right\rfloor  \right) \right\} \,.
	\end{split}
\end{equation}
We arrive at the following matrix factorization
\begin{equation}\label{eq:frakT*B}
	\begin{tikzcd}[scale cd=0.8]
		\T_b^\kappa * B_{c} : \bigoplus_{k\,\text{odd}}\bigwedge^{k} V_{T\otimes B} \otimes R\{q^k-d\kappa,q_R^k+2\kappa \}
		\rar[shift left]{\bar{\delta} + \bar{\sigma}} 
		& \lar[shift left]{\bar{\delta} + \bar{\sigma}} \bigoplus_{k\,\text{even}}\bigwedge^{k} V_{T\otimes B}\otimes R\{q^k-d\kappa,q_R^k+2\kappa \}.
	\end{tikzcd}
\end{equation}
Here, $R=\CC[X_1,\ldots,X_N]/(G(X_i))$ and $q^k$ and $q^k_R$ are the charges given in equation~(\ref{eq:qqr}).
The maps $\bar{\delta}$ and $\bar{\sigma}$ read
\begin{equation}
	\begin{split}
		\bar{\delta}^k &:=\bar{\delta}\Big|_{\bigwedge^{k} V_{T\otimes B} \otimes R\{q^k-d\kappa,q_R^k+2\kappa \}}
		= \begin{cases}
			\sum_{i=1}^N\iota_{\widetilde{e}_i^*}\otimes X_i &\quad \text{for $[k-1] \ne d-1-[c-b]$} \\
			0 &\quad \text{for $[k-1] = d-1-[c-b]$} \\
		\end{cases}\\
		\bar{\sigma}^k &:= \bar{\sigma}\Big|_{\bigwedge^{k} V_{T\otimes B} \otimes R\{q^k-d\kappa,q_R^k+2\kappa \}}=
		\begin{cases}
			0 &\quad \text{for $[k] \ne d-1-[c-b]$} \\
			\sum_{j=1}^N (\widetilde{e_i}\wedge\cdot)\otimes X_j^{d-1} &\quad \text{for $[k] = d-1-[c-b]$} 
		\end{cases}
	\end{split}
\end{equation}
$\T_b^\kappa * B_c$ are now matrix factorizations of $-G(Y_i) + G(Y_i) = 0$, i.e.~honest two-periodic complexes
of $\CC[X_i]/(G(X_i))$-modules. 

The morphisms $\varphi^\kappa = \rho_\T +\vartheta_\T:\; \T_b^\kappa
\rightarrow \T_b^{\kappa+1}$ constructed in section~\ref{sec:Transition} (see also equation (\ref{eq:fermmor})) descend to morphisms 
\begin{equation}
\bar{\varphi}^\kappa=r_{\T}\circ(\varphi^\kappa\otimes\text{id}_{Q_c})\circ r^*_\T: \T_b^\kappa * B_c \rightarrow \T_b^{\kappa+1} * B_c
\end{equation}
of the fusion products. They also decompose as 
$\bar\varphi^\kappa = \bar\rho_\T +\bar\vartheta_\T$, 
where
\begin{equation}
	\bar{\rho }_\T = r_\T \circ (\rho_\T\otimes\id_{Q_c}) \circ r_\T^*\,,\quad\text{and}\quad
	\bar{\vartheta }_\T = r_\T \circ (\vartheta_\T\otimes\id_{Q_c}) \circ r_\T^*\,.
\end{equation}
are given by
\begin{align}
	\bar{\rho }_\T^k &:=\bar{\rho}_\T\Big|_{\bigwedge^{k} V_{T\otimes B} \otimes R\{q^k-d\kappa,q_R^k+2\kappa \}}	
	= \delta_{[d-k], [c-b]}\left(\sum_{i=1}^N\iota_{e_i^*}\otimes X_i\right)   , \\
	\bar{\vartheta }_\T^k &:=\bar{\vartheta}_\T\Big|_{\bigwedge^{k} V_{T\otimes B} \otimes R\{q^k-d\kappa,q_R^k+2\kappa \}}
	= (1-\delta_{[d-k-1], [c-b]})\left(\sum_{i=1}^N (e_i\wedge\cdot)\otimes X_i^{d-1} \right) \,.
\end{align}
These ingredients can be inserted into equation (\ref{eq:TxB-cone}), and one obtains the fusion product as a successive cone
\begin{equation*}
	\begin{split}
		\T_b * B_{c} &= \cone \left( \bar{\varphi}^0 : \T_b^0 * B_{c} \rightarrow \cone\left(
		\bar{\varphi}^1 : \T_b^1 * B_{c} \rightarrow \cone\left(\ldots\right) \right) \right) ,
	\end{split}
\end{equation*}
which is a complex of $R=\CC[X_i]\big/( G(X_i))$-modules. Following \cite{HHP_08}, this complex can 
be interpreted as a complex of coherent sheaves on the projective hypersurface $M = \{ G(X_i) = 0 \} \subset
\mathbb{P}^{N-1}$ by replacing free modules $R\{q,q_R\}$ with $\O_M(-q)[-q_R]$, where $\O_M$ is the structure sheaf on  $M = \{ G(X_i) = 0 \} \subset
\mathbb{P}^{N-1}$.\footnote{$(\cdot)$ denotes the twisting of the sheaf and $[\cdot]$ the shift of complexes.}

\subsubsection{Results}

We are now ready to concretely compute the fusion of the transition defects $\T_b$ with the LG branes $B_c$.
Here, we assume $d\geq N$, so that the geometric phase lies in the IR of the theory.\footnote{This assumption is made purely for the interpretation of the results. The construction of the defects $\T_b$ is general and the calculation of their fusion with $B_c$ can also be carried out for $d<N$.}

For simplicity and in order to compare our results with the example considered in \cite{HHP_08}, we
first look at the case $N=3$ before generalising to arbitrary $N$.

\subsubsection*{Example $N=3$}

In this case, unfolding the $\T_b^\kappa * B_{c}$ with respect to $U(1)_R$-charge gives, up to a shift of gauge charges, four different complexes depending on the value of $[c-b]$.  For $0\le \{c-b\}_d< d-N$ we get a complex of the form 
\begin{equation}\label{eq:N=3-complexes-parts-1}
		\begin{tikzcd}
			R_{2(\kappa+l)-3} \rar{\bar{\delta }^3}
			&	R_{2(\kappa +l) -2}^{\oplus 3} \rar{\bar{\delta }^2} 
			& R_{2(\kappa +l) -1}^{\oplus 3} \rar{\bar{\delta }^1}
			& \underline{R_{2\left( \kappa + l \right)}} 
		\end{tikzcd},
\end{equation}
where the underline denotes position $0$ in the complex, the subscripts denote the
$U(1)_R$-charge with $l = \left\lfloor \frac{c-b }{d} \right\rfloor$. 
The term at position $-k$ in the complex corrsponds to the submodule $\bigwedge V_{T\otimes B}\otimes R\{q^k-d\kappa,q^k_R+2\kappa\}$ in the representation (\ref{eq:frakT*B}) of $\T_b^\kappa*B_c$. 

To simplify notation
we dropped the gauge charges, which can be read off from equation~\eqref{eq:frakT*B-charges}. We will
reintroduce them later when collecting our results. The complexes corresponding to the remaining
possible values of $[c-b]$ are given by \\

\noindent $[c-b] = [d-3]$:
\begin{equation}
		\begin{tikzcd}[ampersand replacement=\&]
			R_{2(\kappa + l) -2}^{\oplus 3} \rar{\bar{\delta }^2} \ar{dr}{\bar{\sigma }^2} 
			\& R_{2(\kappa + l) -1}^{\oplus 3} \rar{\bar{\delta }^1} \ar[phantom]{d}[marking]{\oplus}
			\& \underline{R_{2(\kappa + l)}}
			\\
			\& R_{2(\kappa + l)-1}
		\end{tikzcd}
\end{equation}
$[c-b] = [d-2]$:
\begin{equation}
		\begin{tikzcd}[ampersand replacement=\&]
			R_{2(\kappa + l) -1}^{\oplus 3} \rar{\bar{\delta }^1} \ar{dr}{\bar{\sigma }^1} \dar[phantom,marking]{\oplus}
			\& R_{2(\kappa + l)} \dar[phantom,marking]{\oplus}
			\\
			R_{2(\kappa + l)-1} \rar{\bar{\delta }^3}
			\&	\underline{R_{2(\kappa + l) }^{\oplus 3}}
		\end{tikzcd}
\end{equation}
$[c-b] = [d-1]$:
\begin{equation}\label{eq:exunwrap}
		\begin{tikzcd}[ampersand replacement=\&]
			\& R_{2(\kappa + l)} \ar{dr}{\bar{\sigma }^0} \dar[phantom,marking]{\oplus}
			\\
			R_{2(\kappa + l)-2} \rar{\bar{\delta }^3}
			\& \underline{R_{2(\kappa + l) -2}^{\oplus 3}} \rar{\bar{\delta }^2}  
			\& R_{2(\kappa + l) -1}^{\oplus 3}  
		\end{tikzcd}
\end{equation}
To get the total complex corresponding to $\T_b*B_{c}$, we bind these individual complexes together using the maps $\bar{\rho }^k$ and $\bar{\vartheta}^k$ derived above. 
Additionally, we want to write it as a complex of coherent sheaves, which is achieved by
sheafification (\cite{BJR-Monodromies, HHP_08, Eisenbud}). As mentioned above, this means we replace
the module $R_{q_R}\{q\}$ with the sheaf $\O(-q)[-{q_R}]$. ($q$ and ${q_R}$ are the $U(1)$- and $R$-charge, respectively.) 
\\

As a concrete example we consider the case 
$$
b=-d+2\,,\qquad
c=[1]\,.
$$
In this case $[c-b]=[d-1]$, so the unwrapped form of $\T_b^\kappa * B_c$ is given by (\ref{eq:exunwrap}). 
Fusion of $\T_b$ with $B_c$ yields the following complex of coherent sheaves on $M$:
\begin{equation*}
		\begin{tikzcd}[ampersand replacement=\&, column sep=small, row sep=small, cramped]
			\& \O_M(-1)_{0}  
			\ar{dr}[inner sep=2pt]{\bar{\sigma }^0} \dar[phantom,marking]{\oplus}
			\\
			\O_M(d-4)_{-1}  
			\rar{\bar{\delta }^3}
			\& \underline{\O_M(d-3)_{0}^{\oplus 3}} 
			\rar{\bar{\delta }^2}  
			\ar[blue]{dr}[inner sep=2pt]{\bar{\vartheta}^2}
			\& \O_M(d-2)_{1}^{\oplus 3} 
			\rar[blue]{\bar{\rho }^1}
			\ar[blue]{dr}[inner sep=2pt]{\bar{\vartheta}^1}
			\dar[phantom,marking]{\oplus}
			\& \O_M(d-1)_{2} 
			\ar{dr}{\bar{\sigma }^0} \dar[phantom,marking]{\oplus}
			\\
			\&
			\&
			\O_M(2d-4)_{1} 
			\rar{\bar{\delta }^3}
			\& \O_M(2d-3)_{2}^{\oplus 3} 
			\rar{\bar{\delta }^2}  
			\ar[blue]{dr}[inner sep=2pt]{\bar{\vartheta}^2}
			\& \O_M(2d-2)_{3}^{\oplus 3} 
			\rar[blue]{\bar{\rho }^1}
			\ar[blue]{dr}[inner sep=2pt]{\bar{\vartheta}^1}
			\& \cdots 
			\\
			\& \& \& \& 
			\phantom{\cdots}\cdots \phantom{\cdots}
			\& \cdots 
		\end{tikzcd}
\end{equation*}
The maps coloured in blue arise from the cone construction and the remaining connected blocks come from the
individual components $\T_{b} * B_{c}$. 

Note that all the horizontal maps are given by $\delta=\sum_i \iota_{\widetilde{e}_i^*}\otimes X_i $ and all
the diagonal maps correspond to $\sigma=\sum_i (\widetilde{e}_i \wedge \cdot)\otimes  X_i^{d-1}$. We see that in this case 
$\T_{b} * B_c$ is given by the bound state of the large volume D-brane $\O_M(-1)_0$ and infinitely many copies of the D-brane corresponding to the complex 
\begin{equation}\label{eq:TxB-trivial-line}
	\begin{tikzcd}
		\O_M(q-3)_{{q_R}-3}  
		\rar{\delta}
		& \O_M(q-2)_{{q_R}-2}^{\oplus 3} 
		\rar{\delta}  
		& \O_M(q-1)_{{q_R}-1}^{\oplus 3} 
		\rar{\delta}  
		& \O_M(q)_{{q_R}} .
	\end{tikzcd}
\end{equation}
As shown in \cite{HHP_08}, this complex corresponds to the trivial D-brane
in the geometric phase, i.e.~it is quasi-isomorphic to the $0$-complex. 
This means we get 
\begin{equation}
	\cone( \O_M(-1)_0[-1] \rightarrow 0 ) \cong \O_M(-1)_0
\end{equation}
for the complex of coherent sheaves associated to $\T_{b} * B_{c}$.
For the case $d=N$ this result agrees with the one of \cite{HHP_08}.\footnote{In \cite{HHP_08} only the Calabi-Yau case, i.e.~$d=N$ is treated.}

The above discussion easily carries over to arbitrary values for $b$ and $c$. Again one can eleminate subcomplexes of the form~\eqref{eq:TxB-trivial-line} corresponding to trivial branes, and obtains 
\begin{equation}
	\begin{aligned}
		0 \quad &\text{for $0 \leq \{c-b\}_d < d-3$}\\
		\O_M^{\oplus 3}(\begin{smallmatrix} -b-d+1 \end{smallmatrix}) \rightarrow \O_M^{\oplus 3}(\begin{smallmatrix} -b-d+2 \end{smallmatrix}) \rightarrow \underline{\O_M(\begin{smallmatrix} -b-d+3 \end{smallmatrix})} \quad &\text{for $[c-b] = [d-3]$}\\
		\O_M^{\oplus 3}(\begin{smallmatrix} -b-d+1 \end{smallmatrix}) \rightarrow \underline{\O_M(\begin{smallmatrix} -b-d+2 \end{smallmatrix})} \quad &\text{for $[c-b] = [d-2]$}\\
		\underline{\O_M(\begin{smallmatrix} -b-d+1 \end{smallmatrix})} \quad &\text{for $[c-b] = [d-1]$}
	\end{aligned}
\end{equation}
as complexes of coherent sheaves associated to $\T_{b} * B_{c}$.

\subsubsection*{General results}

The generalization to arbitrary $N$ is straight-forward, following the same steps as above.
Unfolding the individual products $\T_b^\kappa * B_c$ with respect to $U(1)_R$-charge now gives
complexes of the form 
\begin{equation*}
	\begin{tikzcd}[column sep=small, row sep=small, cramped]
		& & & & R^{\left( \begin{smallmatrix}
			N \\ d-1-a
		\end{smallmatrix} \right)} \rar{\delta} 
		\ar{dr}{\sigma}
		\dar[phantom,marking]{\oplus}
		& \cdots \rar{\delta}
		\dar[phantom,marking]{\oplus}
		& R^{\left( \begin{smallmatrix}
			N \\ 1 
		\end{smallmatrix}\right)} \rar{\delta} 
		& R
		\\
		& & R^{\left( \begin{smallmatrix}
			N \\ 2d-1-a
		\end{smallmatrix} \right)} \rar{\delta} 
		\ar{dr}{\sigma}
		\dar[phantom,marking]{\oplus}
		& \cdots \rar{\delta}
		\dar[phantom,marking]{\oplus}
		& R^{\left( \begin{smallmatrix}
			N \\ d-a + 1 
		\end{smallmatrix}\right)} \rar{\delta} 
		& R^{\left( \begin{smallmatrix}
			N \\ d-a 
		\end{smallmatrix}\right)} 
		\\
		& 
		& \cdots \rar{\delta} \ar[phantom]{dl}[marking]{\iddots}
		& R^{\left( \begin{smallmatrix}
			N \\ 2d-a 
		\end{smallmatrix}\right)} 
		\\
		R^{\left( \begin{smallmatrix}
			N \\ N
		\end{smallmatrix}\right)} \rar{\delta} & \cdots 
	\end{tikzcd}
\end{equation*}
where $a:=\{c-b\}_d$ and the $U(1)$ and $U(1)_R$-charges can be obtained from \eqref{eq:frakT*B-charges}.
Following the same steps as before, i.e.~binding these complexes together using $\bar{\rho}$ and $\bar{\vartheta}$ as before, replacing modules with sheaves and splitting off infinitely many trivial subcomplexes one arrives at the finite complexes associated to the fusion $\T_b*B_c$. In order to systematically write them down, 
we note that the sheaves 
\begin{equation}
	\Omega_M^t (q) := \Omega_{\mathbb{P}^{N-1}}^t (q)\big|_M = \textstyle\bigwedge^t T^* \mathbb{P}^{N-1} \otimes
	\O_M(q).
\end{equation}
are quasi-isomorphic to the complexes 
\begin{equation}
	\begin{tikzcd}[column sep=small, row sep=small, cramped]
		\Omega_M^t(q) [t] \cong \O_M^{\left( \begin{smallmatrix}
			N \\ t
		\end{smallmatrix} \right)} \rar{\delta} 
		& \cdots \rar{\delta}
		& \O_M^{\left( \begin{smallmatrix}
			N \\ 1 
		\end{smallmatrix}\right)} \rar{\delta} 
		& \underline{\O_M(q)}
	\end{tikzcd}.
\end{equation}
Setting $\Omega^t := \Omega_M^t(t-(b+d-1)) [t]$ then, with  
$$
k:=\left\lfloor\frac{N}{d}\right\rfloor\,,\quad
\text{and}\quad a:=\{c-b\}_d\,,
$$
one can write the finite complex of coherent sheaves associated to $\T_b*B_c$ as 
\begin{equation}
	\cone \big(\Omega^{d-1-a} \rightarrow \cone \big(\Omega^{2d-1-a}[-2] \rightarrow
	\cone \big( \cdots \rightarrow \Omega^{(k+1)d -1 -a}[-2k] \big) \big) \big) .
\end{equation}
The maps binding the complexes $\Omega^t$ together are given by $\sigma$ above.

For $d \ge N$ this reduces to:
\begin{equation}
	\begin{aligned}
		0 \quad &\text{for $0 \leq \{c-b\}_d < d-N$}\\
		\Omega^{N-1}_{M}(-b - d + N  )\{N-1\} \quad &\text{for $[c-b] = [d-N]$}\\
		\Omega^{N-2}_{M}(-b- d + N-1))\{N-2\} \quad &\text{for $[c-b] = [d-N+1]$}\\
		\vdots \\
		\Omega^{1}_{M}(-b-d+2) \{1\} \quad &\text{for $[c-b] = [d-2]$}\\
		\Omega^{0}_{M}(-b-d+1)  \quad &\text{for $[c-b] = [d-1]$} .
	\end{aligned}
\end{equation}
For the case $d=N$ this agrees with the results obtained in 
 \cite{HHP_08} for the transport of the  LG branes $B_c$ into the geometric phase.

In the case $d>N$ we observe that a certain set of D-branes vanishes when fused with $\T_b$. This is of course expected. After all, in this case the transition between LG and geometric phase corresponds to a relevant RG flow, under which certain vacua and with it some D-branes decouple from the theory. 
Our computation specifies which D-branes decouple under the flow. 
Using the terminology of  \cite{hori_exact_2013,RlFC_brane-transport_18}, the D-branes are transported through a ``large window" (determined by $b$) and those with charges contained in a ``small window" are transported to the new conformal fixed point. The others decouple.

\section{Conclusions}

In this paper, we construct defects describing the transition between Landau-Ginzburg and geometric phases of abelian gauged linear sigma models. As it turns out, the construction required certain choices, which 
precisely match with the possible (homotopy classes of) paths between the phases. On the level of D-branes, the defects act by means of fusion, giving rise to functors from the category of D-branes in the Landau-Ginzburg phase (category of equivariant matrix factorizations) to the category of D-branes in the geometric phase (derived category of coherent sheaves on the target space). In a class of examples, we explicitly compute the action of the transition defects on D-branes and find that the results agree with the behavior of LG branes under the transition to the geometric phase studied by other methods in \cite{HHP_08,hori_exact_2013,Knapp:2016rec,RlFC_brane-transport_18}.\footnote{In an interesting recent paper \cite{Galakhov:2021omc} the topic of D-brane transport using defects has been discussed from a different point of view using the approach put forward in \cite{Gaiotto:2015aoa}.}

Our discussion was restricted to gauged linear sigma models with $U(1)$ gauge symmetry, exemplifying the general strategy. We expect that our arguments carry over to abelian gauge symmetries of higher rank in a straight forward way and hope to come back to this in future work.

In this paper, we deal with defects from LG  to  geometric phases of GLSMs. Of course, one could equally well construct analogous defects going in the opposite direction. These are also hybrids between matrix factorizations  and  complexes, and they also factorize over the GLSM. The factors in this case, however, are defects 
embedding the geometric phase into the GLSM and defects pushing down the GLSM  to the LG phase.
 Indeed, the composition of the two types of defects would lead to defects describing monodromies, and in particular should reproduce the monodromy defects derived in \cite{BJR-Monodromies}.

\section*{Acknowledgements}

IB is supported by the Deutsche Forschungsgemeinschaft (DFG, German Research Foundation) under Germany's Excellence Strategy -- EXC-2094 -- 390783311 and the DFG grant  ID 17448. DR is supported by the Heidelberg Institute for Theoretical Studies. DR thanks the MSRI in Berkeley for its hospitality, where part of this work was done. (Research at MSRI is partly supported by the NSF under Grant No. DMS-1440140.)

\appendix

\section{Reduction to finite rank: the isomorphisms $r$ and $r^*$}\label{sec:app}

In this appendix we will show that the maps $r$ and $r^*$ defined in (\ref{eq:defr}), respectively (\ref{eq:defr*}) are  isomorphisms between the unreduced and reduced tensor product matrix factorizations $T_b\otimes B_c$ and $(T_b\otimes B_c)^\text{red}$. In fact, 
using the modules associated to these matrix factorizations we already established in section~\ref{sec:GLSM-action-on-D-branes} that the two matrix factorizations are isomorphic. 
It is not difficult to check that the map $r$ is indeed a morphism of the matrix factorizations and that it descends to the isomorphism of the respective modules. Therefore, $r:T_b\otimes B_c\rightarrow (T_b\otimes B_c)^\text{red}$ is an isomorphism. 

Moreover, just using the definitions 
 (\ref{eq:defr}) and (\ref{eq:defr*}) of $r$ and $r^*$ it is easy to deduce that $r\circ r^*=\text{id}_{(T_b\otimes B_c)^\text{red}}$. The only part which is more involved is to show that the map $r^*$ is indeed a morphism of matrix factorizations, i.e.
 \begin{equation}\label{eq:appid1}
 (\delta_T+\sigma_T+\delta_B+\sigma_B)\circ r^*=
 r^*\circ(\delta_\text{red}+\sigma_\text{red})\,.
 \end{equation}
One can check this identity in a straight-forward manner. However the calculation simplifies dramatically in a different basis. Let $K$ be the operator counting the degree of forms of $V_T$. Instead of (\ref{eq:appid1}) we will check the equivalent identity
\begin{equation}\label{eq:appid2}
(\widetilde{\delta}_T+\widetilde{\sigma}_T+\delta_B+\sigma_B)\circ \widetilde{r}^*=
 \widetilde{r}^*\circ(\widetilde{\delta}_\text{red}+\widetilde{\sigma}_\text{red})\,,
 \end{equation}
 where
 \begin{align}
 &\widetilde{r}^*=\varepsilon^K\circ r^*\circ \varepsilon^{-K} \quad&\\
 &\widetilde{\delta}_T=\varepsilon^K\circ\delta_T\circ \varepsilon^{-K}
 &&\widetilde{\sigma}_T=\varepsilon^K\circ\sigma_T\circ \varepsilon^{-K}\\
 &\widetilde{\delta}_\text{red}=\varepsilon^K\circ\delta_\text{red}\circ \varepsilon^{-K}
 &&\widetilde{\sigma}_\text{red}=\varepsilon^K\circ\sigma_\text{red}\circ \varepsilon^{-K}
 \end{align}
Note that the dependence on the form degree vanishes in this basis. One obtains
\begin{align}
&\widetilde{\delta}_T=\sum_{i=1}^N\imath_{e_i^*}\otimes (X_i\varepsilon_P-Y_i)\quad &&
\widetilde{\sigma}_T=\sum_{i=1}^N(e_i\wedge\cdot)\otimes(\prod_{a=1}^{d-1}X_i\varepsilon_P-\xi^a Y_i)\\
&\widetilde{\delta}_\text{red}=\sum_{i=1}^N\imath_{e_i^*}\otimes X_i\varepsilon_P &&
\widetilde{\sigma}_\text{red}=\sum_{i=1}^N(e_i\wedge\cdot)\otimes(X_i\varepsilon_P)^{d-1}\,,
\end{align}
where
\begin{equation}
\varepsilon_P=I_P\varepsilon^{-1}=\left(
\begin{array}{cccc}
0&1&&\\
&\ddots&\ddots&\\
&&\ddots&1\\
P&&&0
\end{array}
\right)\,.
\end{equation}
$\widetilde{r}^*$ acts as
\begin{equation}
		r^* ({e}_I \otimes \widetilde{f}_\mu^{b,k})  = 		
		\Bigg[ (e_{i_1}+g_{i_1})\otimes\II\wedge\ldots\wedge (e_{i_k}+g_{i_k})\otimes\II
	\wedge
		\left( \sum_{n=0}^N \frac{(\widetilde{\omega})^{\wedge n}}{n!}
		\right) \Bigg] (1\otimes f_\mu^{b,k}),
\end{equation}
where now
\begin{equation}
	\widetilde{\omega} = \sum_{j=1}^N e_j\wedge g_j \otimes \widetilde{\Lambda}_j
\end{equation}
with
\begin{equation}
	\widetilde{\Lambda}_j = \frac{1}{Y_j }\left( \prod_{a=1}^{d-1} (\varepsilon_PX_j  - \xi^a Y_j ) - (\varepsilon_PX_j)^{d-1} \right).
\end{equation}
We will now check (\ref{eq:appid2}) by applying it on $e_I\otimes f$. The RHS becomes
\begin{equation}\label{eq:apprhs}
\begin{split}
\widetilde{r}^*\circ(\widetilde{\delta}_\text{red}+\widetilde{\sigma}_\text{red})(e_I\otimes \widetilde{f}_\mu^{b,k})
=&\sum_{1\leq l\leq k}(-1)^{l-1}(e+g)_{i_1\ldots\widehat{i_l}\ldots i_k}\otimes (\varepsilon_P X_{i_l})\wedge
\sum_{m\geq 0}\frac{\widetilde{\omega}^m}{m!}(1\otimes f_\mu^{b,k})\\
&+\sum_{j=1}^N(e+g)_I\otimes (\varepsilon_PX_j)^{d-1}\wedge \sum_{m\geq 0}\frac{\widetilde{\omega}^m}{m!}(1\otimes f_\mu^{b,k})\,.
\end{split}
\end{equation}
For the LHS one obtains
\begin{equation}
\begin{split}
&(\widetilde{\delta}_T+\widetilde{\sigma}_T+\delta_B+\sigma_B)\circ \widetilde{r}^*(e_I\otimes \widetilde{f}_\mu^{b,k})=\\
&\qquad
\sum_{1\leq l\leq k}(-1)^{l-1}
(e+g)_{i_1\ldots\widehat{i_l}\ldots i_k}\otimes (\varepsilon_P X_{i_l})\wedge\sum_{m\geq 0}\frac{\widetilde{\omega}^m}{m!}(1\otimes f_\mu^{b,k})\\
&\qquad+\sum_{j=1}^N\left(Y_j^{d-1}g_j+\prod_{l=1}^{d-1}(\varepsilon_PX_j-\xi^lY_j)e_j\right)
\wedge (e+g)_I\wedge\sum_{m\geq 0}\frac{\widetilde{\omega}^m}{m!}(1\otimes f_\mu^{b,k})\\
&\qquad+(\widetilde{\delta}_T+\delta_B)(\widetilde{\omega})\wedge(e+g)_I\wedge\sum_{m\geq 0}\frac{\widetilde{\omega}^m}{m!}(1\otimes f_\mu^{b,k})\,.
\end{split}
\end{equation}
Now, $(\widetilde{\delta}_T+\delta_B)(\widetilde{\omega})$ calculates to
\begin{equation}
\begin{split}
(\widetilde{\delta}_T+\delta_B)(\widetilde{\omega})=&
-\sum_{i=1}^N e_i\otimes \left(\prod_{a=1}^{d-1}(\varepsilon_PX_i-\xi^aY_i)-(\varepsilon_PX_i)^{d-1}
\right)\\
&+\sum_{i=1}^Ng_i\otimes\left(-Y_i^{d-1}+(\varepsilon_P X_i)^{d-1}
\right)\,,
\end{split}
\end{equation}
and one arrives at
\begin{equation}
\begin{split}
&(\widetilde{\delta}_T+\widetilde{\sigma}_T+\delta_B+\sigma_B)\circ \widetilde{r}^*(e_I\otimes \widetilde{f}_\mu^{b,k})=\\
&\qquad
\sum_{1\leq l\leq k}(-1)^{l-1}
(e+g)_{i_1\ldots\widehat{i_l}\ldots i_k}\otimes (\varepsilon_P X_{i_l})\wedge\sum_{m\geq 0}\frac{\widetilde{\omega}^m}{m!}(1\otimes f_\mu^{b,k})\\
&\qquad+\sum_{j=1}^N(e+g)_{ji_1\ldots i_k}\otimes (\varepsilon_PX_j)^{d-1}\wedge
\sum_{m\geq 0}\frac{\widetilde{\omega}^m}{m!}(1\otimes f_\mu^{b,k})\,,
\end{split}
\end{equation}
which indeed agrees with (\ref{eq:apprhs}). Thus, we have shown (\ref{eq:appid2}), and therefore that $r^*$ is a morphism of matrix factorizations. 

\bibliographystyle{JHEP}
	\bibliography{Literatur}

\end{document}